\documentclass[reprint, superscriptaddress, amsmath, amssymb, aps, prb]{revtex4-2}
\usepackage{graphicx}% Include figure files
\usepackage{dcolumn}% Ali$G_N$ table columns on decimal point
\usepackage{bm}% bold math
\usepackage{placeins}
\usepackage{caption}

\usepackage[usenames,dvipsnames]{xcolor}
 \usepackage{amsmath}
 \usepackage{amsfonts}
 \usepackage{amssymb}

 \usepackage{bbold}     % for \mathbb{1} as a nice unit matrix symbol
 \usepackage[makeroom]{cancel}  % crossed terms in tables
 \usepackage{multirow}    % merge cells vertically in tables
 \usepackage[normalem]{ulem}        % strike-through text
 \usepackage{array}
 \usepackage{pdfpages}
\makeatletter
 \AtBeginDocument{\let\LS@rot\@undefined}
 \makeatother

\begin{document}

\title{Large zero bias peaks and dips in a four-terminal thin InAs-Al nanowire device}

\author{Huading Song}
 \email{equal contribution}
\affiliation{State Key Laboratory of Low Dimensional Quantum Physics, Department of Physics, Tsinghua University, Beijing 100084, China}
\affiliation{Beijing Academy of Quantum Information Sciences, 100193 Beijing, China}

\author{Zitong Zhang}
 \email{equal contribution}
\affiliation{State Key Laboratory of Low Dimensional Quantum Physics, Department of Physics, Tsinghua University, Beijing 100084, China}

\author{Dong Pan}
 \email{equal contribution}
\affiliation{Beijing Academy of Quantum Information Sciences, 100193 Beijing, China}
\affiliation{State Key Laboratory of Superlattices and Microstructures, Institute of Semiconductors, Chinese Academy of Sciences, P. O. Box 912, Beijing 100083, China}

\author{Donghao Liu}
\affiliation{State Key Laboratory of Low Dimensional Quantum Physics, Department of Physics, Tsinghua University, Beijing 100084, China}

\author{Zhaoyu Wang}
\affiliation{State Key Laboratory of Low Dimensional Quantum Physics, Department of Physics, Tsinghua University, Beijing 100084, China}

\author{Zhan Cao}
\affiliation{Beijing Academy of Quantum Information Sciences, 100193 Beijing, China}

\author{Lei Liu}
\affiliation{State Key Laboratory of Superlattices and Microstructures, Institute of Semiconductors, Chinese Academy of Sciences, P. O. Box 912, Beijing 100083, China}

\author{Lianjun Wen}
\affiliation{State Key Laboratory of Superlattices and Microstructures, Institute of Semiconductors, Chinese Academy of Sciences, P. O. Box 912, Beijing 100083, China}

\author{Dunyuan Liao}
\affiliation{State Key Laboratory of Superlattices and Microstructures, Institute of Semiconductors, Chinese Academy of Sciences, P. O. Box 912, Beijing 100083, China}

\author{Ran Zhuo}
\affiliation{State Key Laboratory of Superlattices and Microstructures, Institute of Semiconductors, Chinese Academy of Sciences, P. O. Box 912, Beijing 100083, China}

\author{Dong E Liu}
\affiliation{State Key Laboratory of Low Dimensional Quantum Physics, Department of Physics, Tsinghua University, Beijing 100084, China}
\affiliation{Beijing Academy of Quantum Information Sciences, 100193 Beijing, China}
\affiliation{Frontier Science Center for Quantum Information, 100084 Beijing, China}

\author{Runan Shang}
 \email{shangrn@baqis.ac.cn}
\affiliation{Beijing Academy of Quantum Information Sciences, 100193 Beijing, China}

\author{Jianhua Zhao}
 \email{jhzhao@semi.ac.cn}
\affiliation{Beijing Academy of Quantum Information Sciences, 100193 Beijing, China}
\affiliation{State Key Laboratory of Superlattices and Microstructures, Institute of Semiconductors, Chinese Academy of Sciences, P. O. Box 912, Beijing 100083, China}

\author{Hao Zhang}
\email{hzquantum@mail.tsinghua.edu.cn}
\affiliation{State Key Laboratory of Low Dimensional Quantum Physics, Department of Physics, Tsinghua University, Beijing 100084, China}
\affiliation{Beijing Academy of Quantum Information Sciences, 100193 Beijing, China}
\affiliation{Frontier Science Center for Quantum Information, 100084 Beijing, China}

%\date{\today}

\begin{abstract}
We report electron transport studies of a thin InAs-Al hybrid semiconductor-superconductor nanowire device using a four-terminal design. Compared to previous works, thinner InAs nanowire (diameter less than 40 nm) is expected to reach fewer sub-band regime. The four-terminal device design excludes electrode contact resistance, an unknown value which has inevitably affected previously reported device conductance. Using tunneling spectroscopy, we find large zero-bias peaks (ZBPs) in differential conductance on the order of $2e^2/h$. Investigating the ZBP evolution by sweeping various gate voltages and magnetic field, we find a transition between a zero-bias peak and a zero-bias dip while the zero-bias conductance sticks close to $2e^2/h$. We discuss a topologically trivial interpretation involving disorder, smooth potential variation and quasi-Majorana zero modes. 
\end{abstract}

\maketitle

\captionsetup[figure]{justification=raggedright}
%\tableofcontents

\section{Introduction}

The decade-long hunting of Majorana zero modes (MZMs) \cite{ReadGreen, Kitaev} in hybrid semiconductor-superconductor nanowires is guided by a simple and elegant theory in 2010 \cite{Lutchyn2010, Oreg2010}. This theory model requires four basic ingredients: a one-dimensional electron system, strong spin-orbit interaction, s-wave superconducting pairing and Zeeman energy. InAs and InSb semiconductor nanowires coupled to a superconductor are the most exhaustively studied material systems to engineer these four ingredients into a single device, aiming for the realization of MZMs \cite{Prada2020, Lutchyn2018Review}. Indeed, every single ingredient could be directly or indirectly probed by electron transport experiments and thus confirmed to be present in those devices. For example, (quasi-) one dimensionality could be revealed by the observation of quantum point contact (QPC)-like quantized conductance plateaus \cite{Zhang2017Ballistic, Kammhuber2016}, a hallmark of ballistic one dimensional electron system; induced superconductivity could be probed by tunneling conductance which resolves a hard induced superconducting gap \cite{Chang2015, Krogstrup2015, Gul2017, Gazibegovic2017}; Zeeman energy is related to various estimations of effective g-factors \cite{CPH2018effectiveg, Michiel2018}; spin-orbit coupling could be indirectly probed by anti-crossings of Andreev levels as well as the anisotropic closing of the superconducting gap for different magnetic field directions \cite{Michiel2018, Jouri2019}. Majorana theory further predicts a quantized tunneling zero-bias conductance peak \cite{DasSarma2001, Law2009, Flensberg2010, Wimmer2011QPC}. Initial experiments have measured various zero-bias peaks (ZBPs) in hybrid nanowires but with a small peak height \cite{Mourik, Deng2012, Das2012, Churchill2013, Finck2013}. These first generation experiments have suffered from finite sub-gap tunneling conductance, the soft gap problem. Theory later suggested that the soft gap is due to disorder at the superconductor-semiconductor interface which leads to spatially non-uniform couplings \cite{Takei2013}. In the following years, much experimental efforts have focused on the optimization of material growth and device control, trying to minimize the disorder level at those interfaces \cite{Gul2018}. Indeed, epitaxial growth of superconductors on semiconductor nanowires show a hard induced superconducting gap and cleaner ZBPs \cite{Deng2016}. Later on, large ZBPs with peak height reaching the order of $2e^2/h$ was also observed \cite{Nichele2017, Zhang2021, JouriThesis}. However, ZBPs with height robustly sticking to the quantized value by varying both magnetic field and gate voltages, as predicted by MZM theory, have not been demonstrated yet. 

Meanwhile, new theory developments have introduced the concept of quasi-Majorana zero modes (quasi-MZM) \cite{TudorQuasi, WimmerQuasi}, a type of zero-energy Andreev bound states (ABS) \cite{Prada2012, Cayao2015, Silvano2014, Liu2017, Loss2018ABS} with a topologically trivial origin. The key idea is that every ABS could be mathematically decomposed into two quasi-MZMs. In some regimes with disorder or smooth potential variation \cite{DasSarma2011, BrouwerSmooth, Prada2012}, the tunneling probe only couples to the first quasi-MZM. The second quasi-MZM, while partially overlapping with the first one in space (thus topologically trivial), has negligible coupling to the first one due to opposite spin. Therefore, with only one quasi-MZM contributing to tunneling, ZBPs can be quantized, mimicking topological MZMs.     

On the device part, previous ZBP experiments used two-terminal device designs where the electrode contact resistance remains unknown. This unknown resistance introduces a systematic uncertainty, and if used as a fitting parameter, non-quantized ZBPs could be fitted to the quantized value \cite{Yu}. Improving device fabrication for lower contact resistance could reduce this uncertainty to some extent. For example, one could estimate the upper bound of this resistance based on QPC plateaus in ballistic devices, or properties of the superconducting gap which should remain as a constant by varying the tunnel barrier height \cite{Zhang2021}. If the contact resistance is under- or overestimated, the gap size, extracted from bias voltage, would vary after subtracting the bias drop shared by this inaccurate contact resistance. In addition, if the contact resistance is not an unkown constant but depends on bias, gate voltage or magentic field, the ZBP height and shape will be affected in a more complex way. To solve these uncertainty problems, here we use a four-terminal device design, the first time for ZBP experiments to the best of our knowledge, aiming for quantized ZBPs. 

Another improvement compared to previous works is the InAs nanowire diameter which is $\sim$ 40 nm or below, much thinner than those commonly used in literatures (typically $\sim$ 100 nm). The motivation is two-fold. First, the simplest MZM model assumes a one-dimensional electron system, i.e. single sub-band occupation. However, previous InAs/InSb devices likely have multiple sub-bands occupied in the nanowire region underneath the superconductor. The top sub-band, which hosts MZMs, usually has a much smaller coupling to the normal probe than the lower sub-bands due to smooth barrier potential \cite{Brouwer2012ZBP}. This small tunnel coupling of MZM leads to a narrow ZBP whose quantized height (at zero temperature) could easily be destroyed by thermal averaging \cite{Loss2013ZBP}. Therefore, fewer or ultimately single sub-band occupation is preferred for observing quantized ZBPs. To reach this regime, we reduce the wire diameter by growth. Though occupation number can not be directly probed, a lower number of occupied sub-bands is expected due to the small diameter which is already comparable to InAs band bending size \cite{LutchynSchrodinger}, and thus enhancing the sub-band energy spacing. The second motivation is material quality. Thick InAs nanowires (diameter larger than 50 nm) often exhibit randomly distributed twin defects and stacking faults \cite{Caroff2009, Shtrikman2009, Pan2014}, uncontrolled sources of disorder. Since disorder is currently the biggest obstacle in Majorana devices \cite{GoodBadUgly, DasSarma2021Disorder, Tudor2021Disorder}, thinner InAs nanowires with pure-phase crystal structure could suppress this type of disorder. We do note that other disorder sources, e.g. InAs/Al surface oxides and gate/dielectric imperfections, are still present which remains as a future task. 

\section{Experiment and Discussion}

Fig. 1a shows a scanning electron micrograph (SEM) of the InAs-Al nanowire device. Growth details can be found in Ref. \cite{Pan2020}. N1, N2 and S1, S2 label the four contact electrodes on the normal part of the nanowire and the superconducting part, respectively. The upper side-gate is labeled as TG (tunnel gate) for tunnel barrier tuning. The lower side-gate and global back gate are labeled as SG and BG, respectively.

\begin{figure}[htb]
\includegraphics[width=\columnwidth]{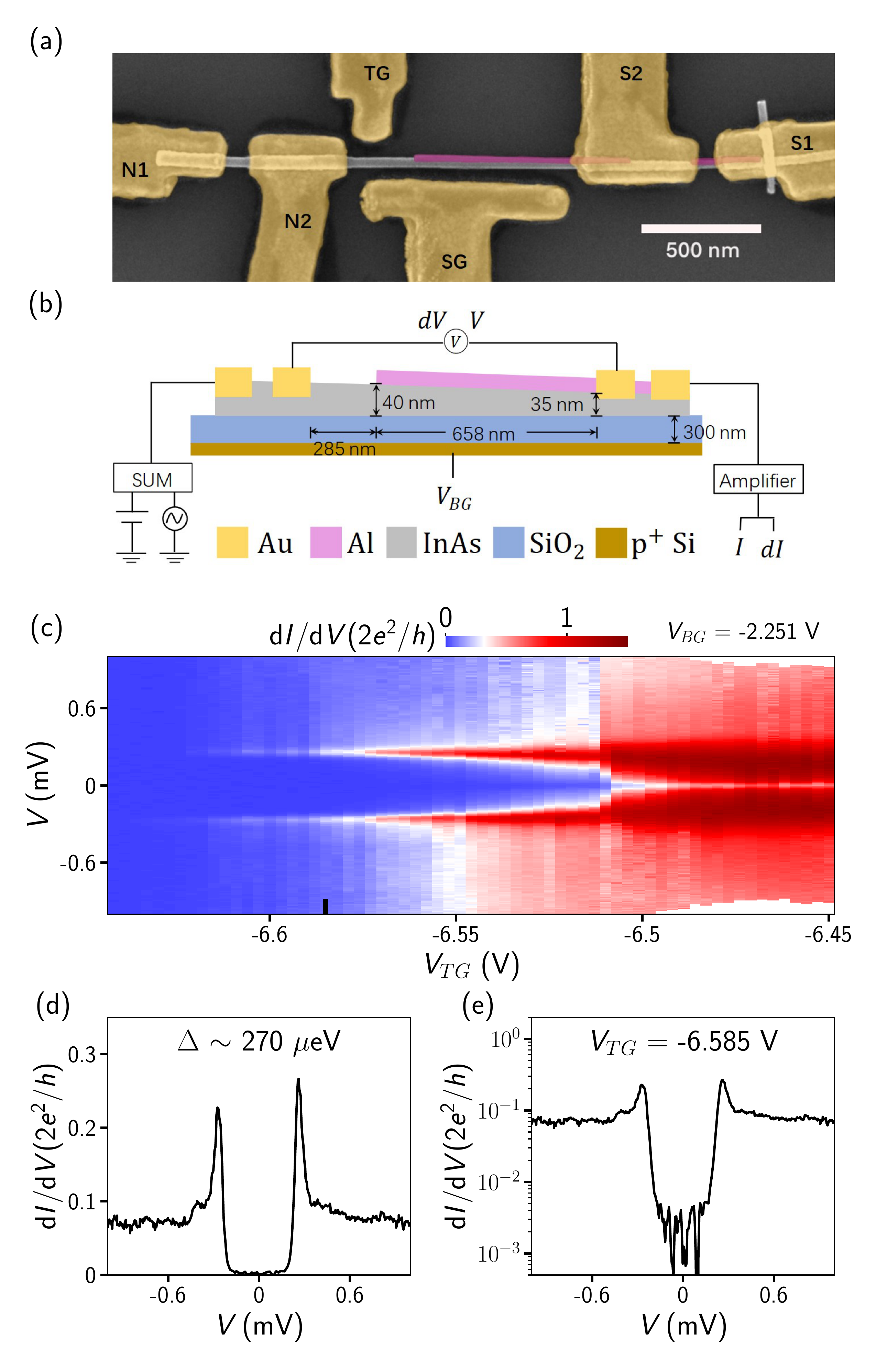}
\centering
\caption{(a) False-color SEM of the device. Four normal electrodes (N1, N2, S1 and S2) and two side gates (TG and SG) are Ti/Au (5/70 nm). The InAs nanowire (gray) is partially covered by a thin Al shell (pink, thickness $\sim$ 7 nm). Part of the Al shell was wet etched for contacts and tunnel barrier. (b) Device and measurement circuit schematic, with dimensions (not in scale) labeled. The substrate (brown) is p+ Si covered by 300 nm thick SiO$_2$ (blue), acting as a global back gate (BG). (c) $dI/dV$ versus bias voltage $V$ and tunnel gate voltage $V_{TG}$ at zero magnetic field. $V_{BG}$ = -2.251 V. (d, e) Vertical line-cut from (c) (black bar) in linear (d) and logarithmic scale (e). }
\label{fig1}
\end{figure}

\begin{figure*}[htb]
\centering
\includegraphics[width=0.8\textwidth]{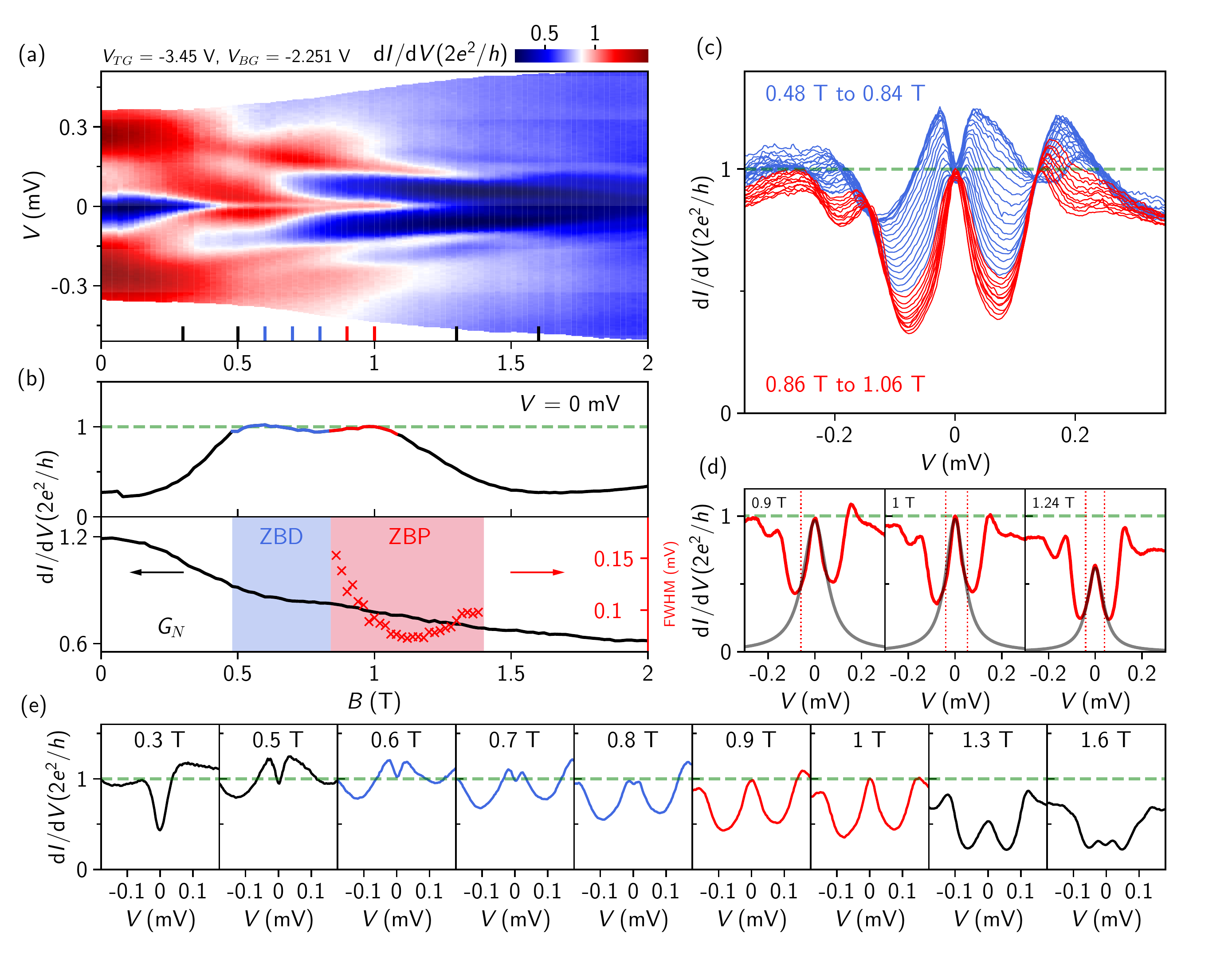}
\caption{(a) $dI/dV$ versus $V$ and $B$ at fixed $V_{TG}$ and $V_{BG}$ (see labeling). $B$ is aligned with the nanowire axis for all measurements. (b) Extracted zero-bias (upper panel) and outside-gap conductance $G_N$ (lower panel). Blue and red backgrounds in the lower panel mark the $B$-ranges of zero-bias dip (ZBD) and ZBP. Blue and red lines in the upper panel mark the $B$-range where the conductance is close to $2e^2/h$. The full width at half maximums (FWHM) of ZBPs are shown as red crosses. (c) $dI/dV$ line-cuts from 0.48 T to 1.06 T, resolving a transition from ZBD (blue) to ZBP (red) near $2e^2/h$. (d) ZBP line-cuts (red) at 0.9, 1 and 1.24 T, together with Lorentzian line shapes (gray) assuming a thermal broadening of 50 mK. Vertical red dashed lines mark the bias positions of ‘half maximum’. (e) Several $dI/dV$ line-cuts ($B$ labeled).}
\label{fig2}
\end{figure*}

Fig. 1b shows the device schematic with dimensions labeled together with a brief measurement circuit (see Supplement Fig. 1 for details). A DC bias voltage and a small AC lock-in excitation are first summed and then applied together to the N1 contact. The resulting current is drained from the S1 contact and measured through a pre-amplifier, as $I$ and $dI$. The voltage drop between N2 and S2 is measured using a voltage meter and another lock-in (synchronized with the first one) to get $V$ and $dV$. Therefore, the differential conductance $dI/dV$ can be directly calculated without subtracting any series resistance (e.g. fridge filters or contacts) as was done before for two-terminal designs. In addition, the bias $V$ can also be directly measured without subtracting the bias drop over series resistance as in the two-terminal case.

Fig. 1c shows $dI/dV$ as a function of $V$ and $V_{TG}$, resolving a hard induced superconducting gap in the tunneling regime (Fig. 1de) where the outside-gap conductance is much smaller than $2e^2/h$. The gap size is $\sim$ 0.27 meV. Fridge base temperature is $\sim$ 20 mK for all measurements. The gap remains hard at finite magnetic field $B$ before its closing, see SFig. 2 $B$-dependence and more gate scans. The super-gate (SG) is not well functional (see SFig. 3 for details), thus fixed at 0 V for all measurements unless specified. The hard gap resolved by tunneling conductance in the gate-$B$ parameter space (where large ZBPs are observed) is an important prerequisite for the search of quantized ZBPs: in soft gap devices \cite{Yu}, Majorana ZBPs are not even expected to be quantized due to severe dissipation broadening \cite{LiuDissipation}.

Next we apply $B$ along the nanowire axis, searching for possible MZM signatures at different gate voltages (see SFig. 4 unsuccessful searches). Fig. 2a shows such a $B$-scan example at a particular gate voltage setting (labeled in the figure) with the zero-bias line-cut and $G_N$ shown in Fig. 2b. As $B$ increases, two broad levels detach from the gap edges and merge at zero energy at $\sim$ 0.5 T. They first form a zero-bias dip (ZBD) which later on evolves into a ZBP. The zero-bias conductance, during this dip-to-peak transition (from 0.48 T to 1.06 T), sticks close to $2e^2/h$. The mean value of the zero-bias conductance within this $B$-range of 0.58 T (blue and red lines in Fig. 2b), is 0.98 with a standard deviation of 0.02, both in unit of $2e^2/h$. All the $dI/dV$ line-cuts within this $B$-range are shown in Fig. 2c (blue for ZBDs and red for ZBPs). The smooth ZBD-ZBP crossover in $B$-scan where the zero-bias conductance sticks close to $2e^2/h$ is a new observation of this paper. 

For $B$ higher than 1.06 T, the ZBP-height quickly decreases away from $2e^2/h$, and finally the peak splits (line-cuts shown in SFig. 5). The lower panel of Fig. 2b marks the full $B$-range of ZBP with a red background, where its difference with the red line in the upper panel indicates the $B$-range of ZBP whose height significantly decreases away from $2e^2/h$. Fig. 2d shows three ZBP line-cuts with the full width at half maximum (FWHM) indicated by the red dashed lines. FWHM is extracted by the bias $V$ where $dI/dV$ is half of its zero-bias conductance. For some cases (e.g. 0.9 T in Fig. 2d), the background conductance in the positive bias region is larger than the ‘half maximum’. FWHM is then taken by doubling the $|V|$ found in the negative bias region. If the background conductance is larger than the ‘half maximum’ for the entire bias range, then no FWHM is extracted. The gray lines are calculated Lorentzian line-shapes of $G_0/(1+(eV/\Gamma)^2)$ after assuming a thermal broadening of 50 mK, showing a rough match with the ZBPs. $G_0$ is the zero-bias conductance while 2$\Gamma/e$ the extracted FWHM, shown as red crosses in Fig. 2b. 

The blue line in Fig. 2b indicates the $B$-range of ZBD with zero-bias conductance sticking close to $2e^2/h$. We note that line-cuts at lower $B$ (e.g. 0 T) also have line-shapes of ‘ZBD’. This dip, a suppression by the superconducting gap, is however different from the ZBDs we quoted in Fig. 2bc, a result of two merging levels.

The outside-gap conductance $G_N$ shows a decreasing trend in Fig. 2b, suggesting a $B$-dependent barrier height since $G_N$ is proportional to the barrier transmission. However, our $G_N$ is extracted by averaging the conductance for the most positive and negative bias voltages available from the data. The available bias range (the color map edge in Fig. 2a) is not too far away from the gap edge, leading to an overestimation of $G_N$, especially at lower $B$. For a more accurate estimation, we use ZBP-width as the indicator of barrier transmission in the $B$-range with no obvious peak splitting. In Fig. 2bc, the FWHM decreases as increasing $B$ (from 0.86 T to 1.06 T), indicating that the barrier height (transmission) possibly increases (decreases). Within this $B$-range, the ZBP-width varies by $\sim$ 50$\%$, much larger than the variation of ZBP-height: $\sim$ 5$\%$ near $2e^2/h$. The relative variation of ZBP-height near $2e^2/h$ being significantly smaller than the relative variation of ZBP-width was used before to argue for a quantized ZBP \cite{Zhang2021}. For higher $B$, the FWHM starts to increase while the ZBP-height shows a continuous decrease. Both are likely due to peak splitting which becomes visible where the splitting is large enough to be resolved, e.g. see the 1.6 T line-cut in Fig. 2e. 

\begin{figure}[htb]
\centering
\includegraphics[width=\columnwidth]{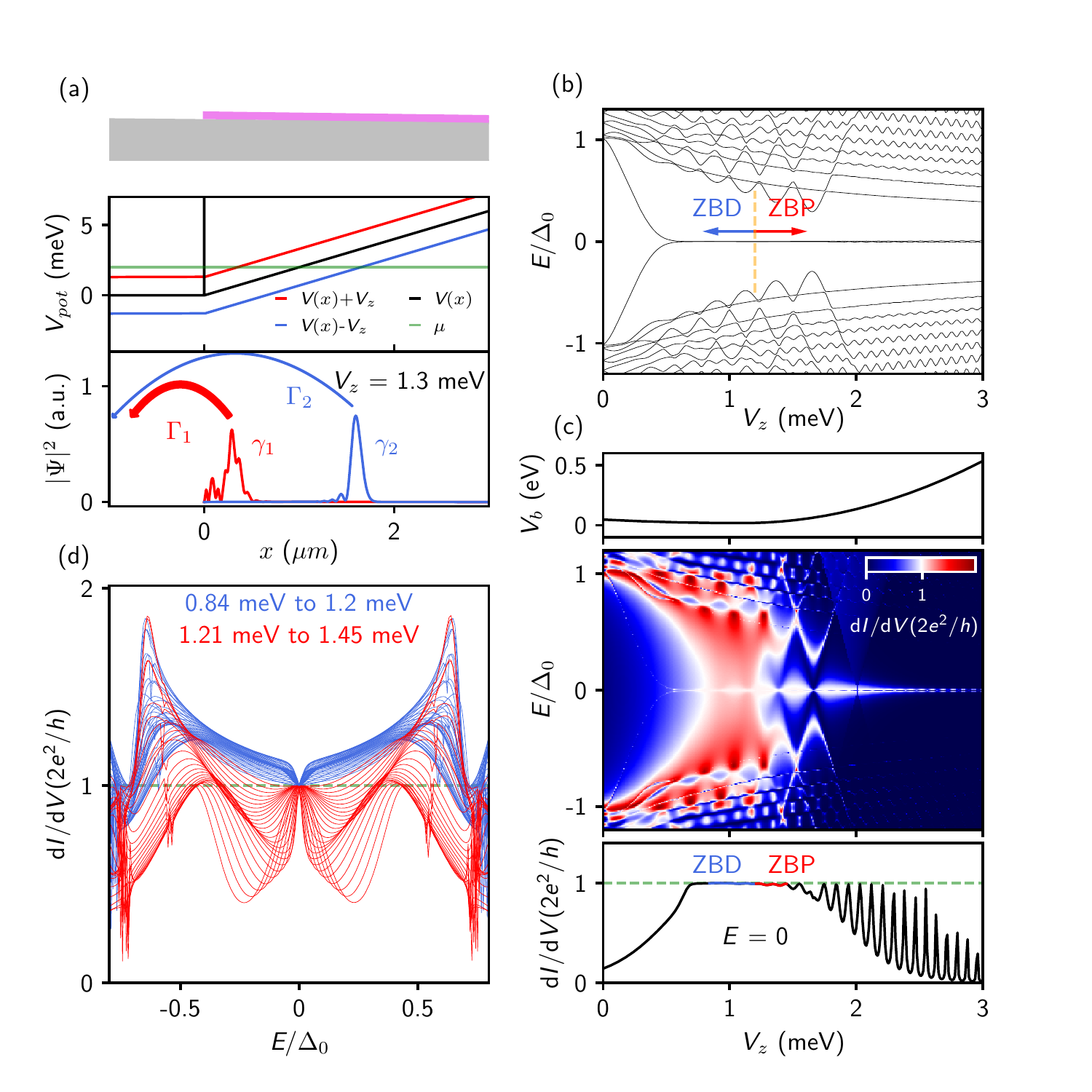}
\caption{Numercial simulation. (a) nanowire schematic (upper), potential landscape (middle) and two quasi-MZM wavefunctions at $V_z$ = 1.3 meV (lower). (b) Energy spectrum. (c) $dI/dV$ of the energy spectrum (middle), zero-energy (bias) line-cut (lower), and $V_z$-dependent barrier height $V_b$ (upper). (d) Line-cuts from (c), quantized ZBD-ZBP transition.}
\label{fig3}
\end{figure}

\begin{figure*}[bt]
\centering
\includegraphics[width=\textwidth]{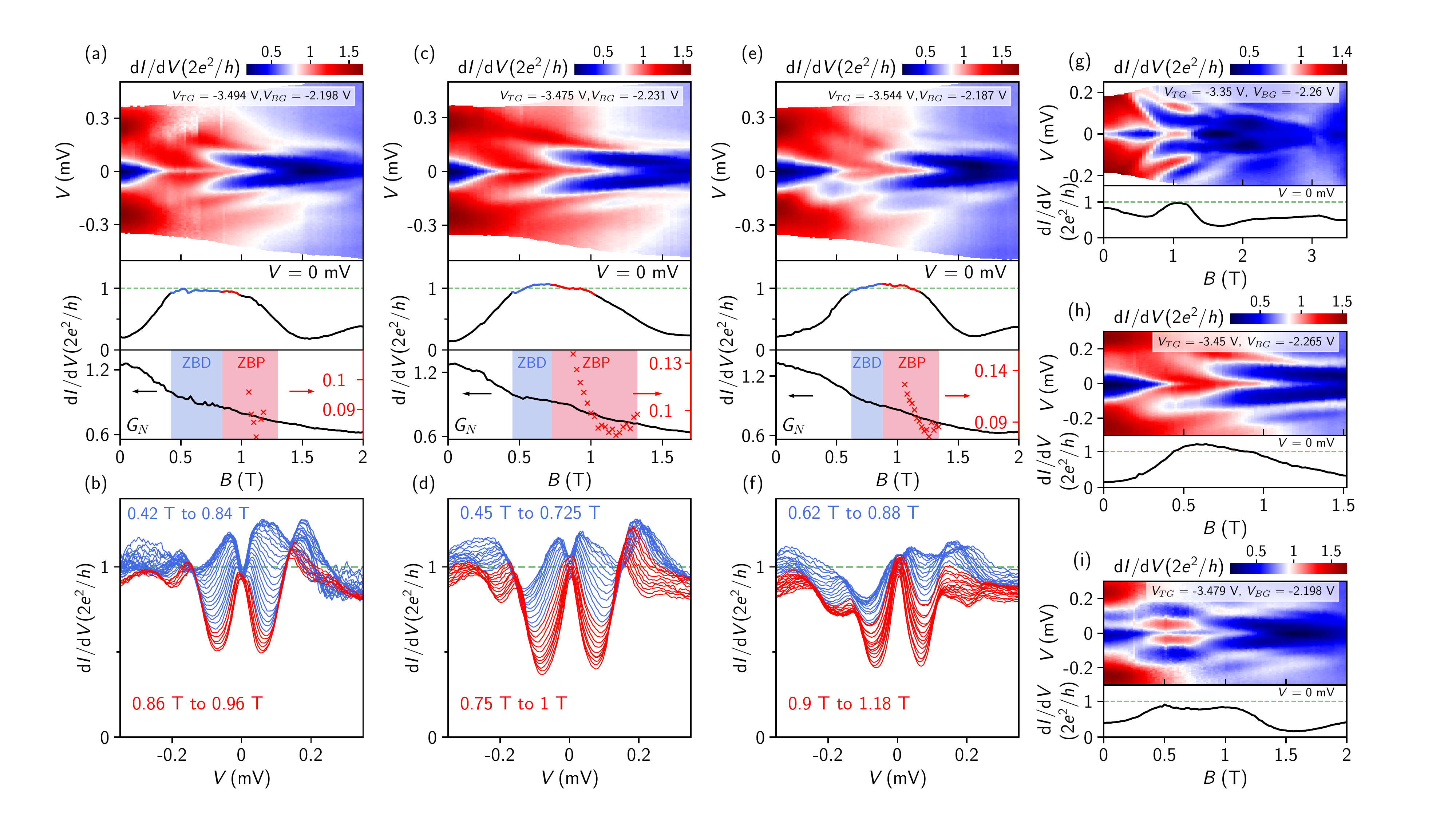}
\caption{(a) $dI/dV$ versus $V$ and $B$ at a different $V_{TG}$ and $V_{BG}$ (see labeling) with the zero-bias line-cut, extracted $G_N$ and FWHM (red crosses) shown in the lower panels. The right y-axis (red) is for FWHM in unit of mV. (b) Line-cuts of ZBD and ZBP near $2e^2/h$ with the $B$-ranges labeled. (c, d) and (e, f), same with (a, b) but different in gate voltage settings (labeled). (g, h, i), Three more $B$-scans with maximum ZBP-heights (g) close to, (h) above and (i) below $2e^2/h$ (gate voltages labeled). Lower panels show zero-bias line-cuts. In (g), the ZBP first splits, then merges back at $B \sim$ 2 T.}
\label{fig4}
\end{figure*}

\begin{figure*}[htb]
\centering
\includegraphics[width=\textwidth]{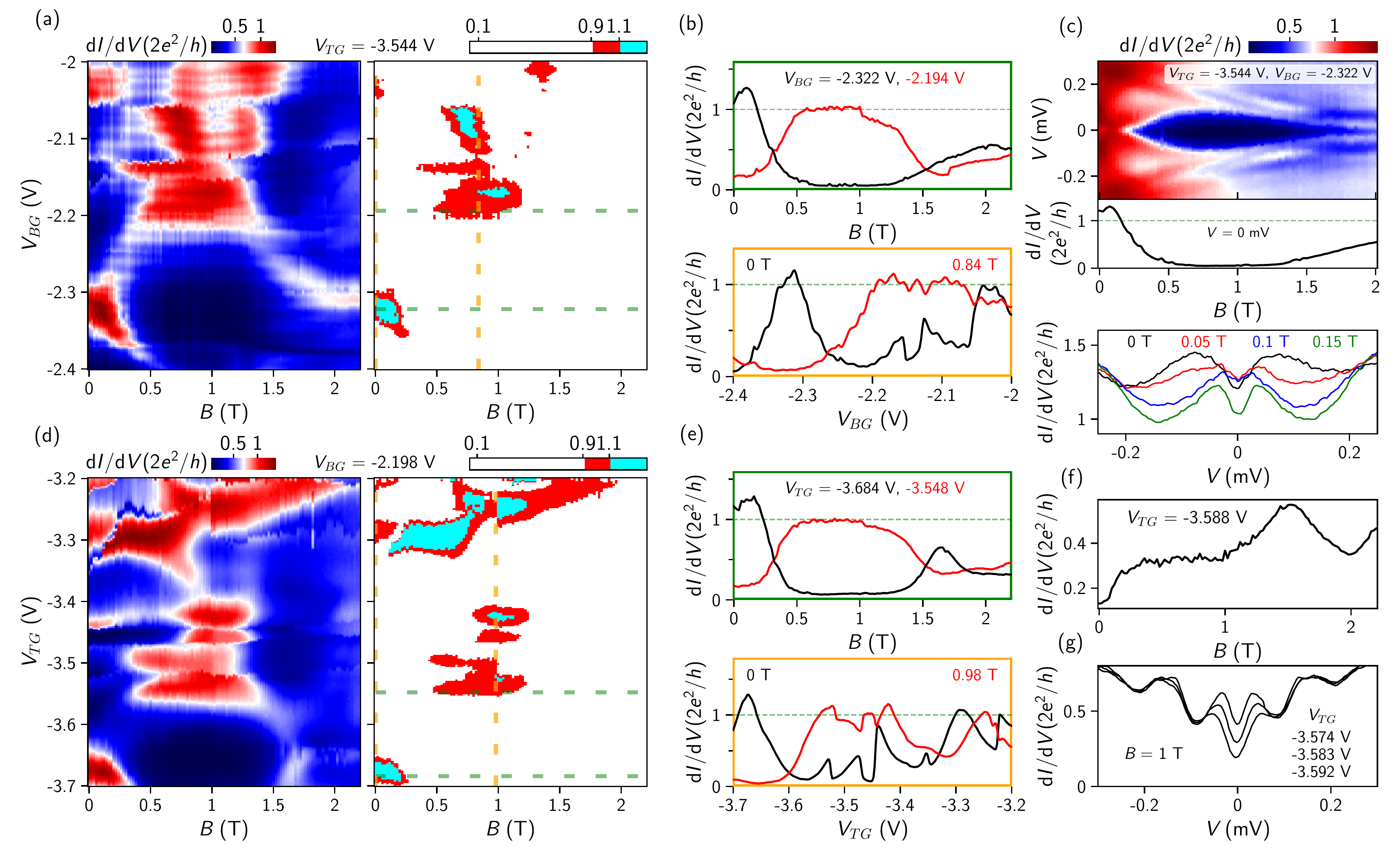}
\caption{(a) $dI/dV$ versus $V_{BG}$ and $B$ (left panel) with $V=0$, $V_{TG}$ = -3.544 V. Right panel: re-plot of the left panel using three colors: white, cyan and red for conductance less than 0.9$\times 2e^2/h$, larger than 1.1$\times 2e^2/h$ and in between. (b) Upper (lower) panel, two horizontal (vertical) line-cuts from (a), indicated by the green (orange) dashed lines, see labeling for $V_{BG}$ ($B$). (c) Upper panel, bias dependence of the black curve in the upper panel of (b). Middle panel, zero-bias line-cut. Lower panel, vertical line-cuts at 0, 0.05, 0.1 and 0.15 T, respectively. (d) $dI/dV$ versus $V_{TG}$ and $B$ (left panel) with $V=0$, $V_{BG}$ = -2.198 V. Right panel: three-color re-plot. (e) Upper (lower) panel, two horizontal (vertical) line-cuts from (d), indicated by the green (orange) dashed lines, see labeling $V_{TG}$ ($B$). (f) Horizontal line-cut from (d) at $V_{TG}$ = -3.588 V. (g) $dI/dV$ line-cuts (from Fig. 6d, the left most part) near the $V_{TG}$ setting of (f) at $B$ = 1 T.}
\label{fig5}
\end{figure*}
             
\begin{figure*}[htb]
\centering
\includegraphics[width=\textwidth]{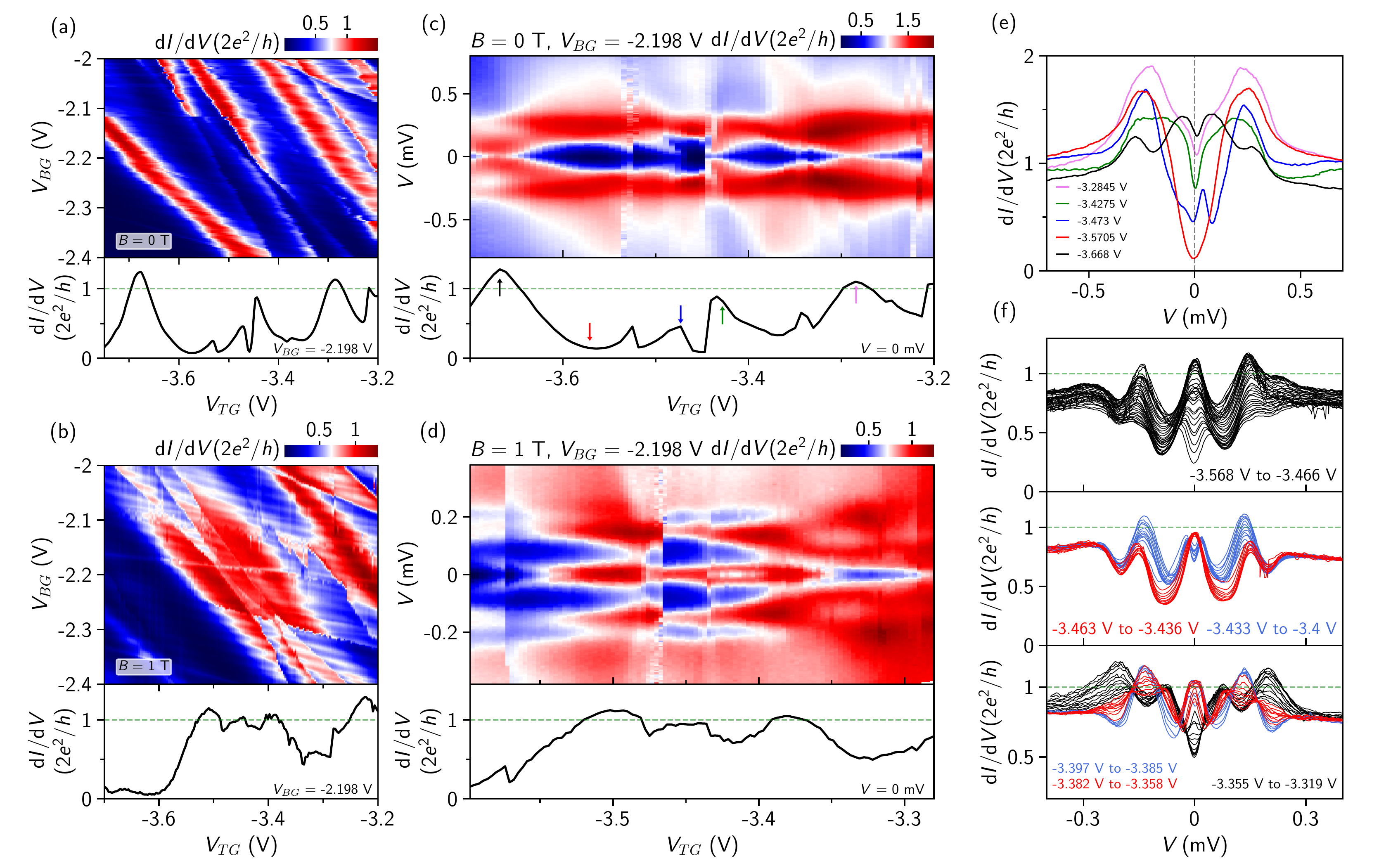}
\caption{(a) $dI/dV$ versus $V_{BG}$ and $V_{TG}$ at $V=0$. $B$ = 0 T. Lower panel, horizontal line-cut at $V_{BG}$ = -2.198 V. (b) Same with (a) except that $B$ = 1 T. (c) Bias dependence of the line-cut in (a). Lower panel, zero-bias line-cut. (d) Same with (b) except that $B$ = 1 T. (e) Several vertical line-cuts from (c) (labeled with colored arrows) showing no obvious ZBPs. (f) Line-cuts from (d) with $V_{TG}$-ranges labeled. For clarity, three panels and colors are used. }
\label{fig6}
\end{figure*}

\begin{figure}[htb]
\centering
\includegraphics[width=\columnwidth]{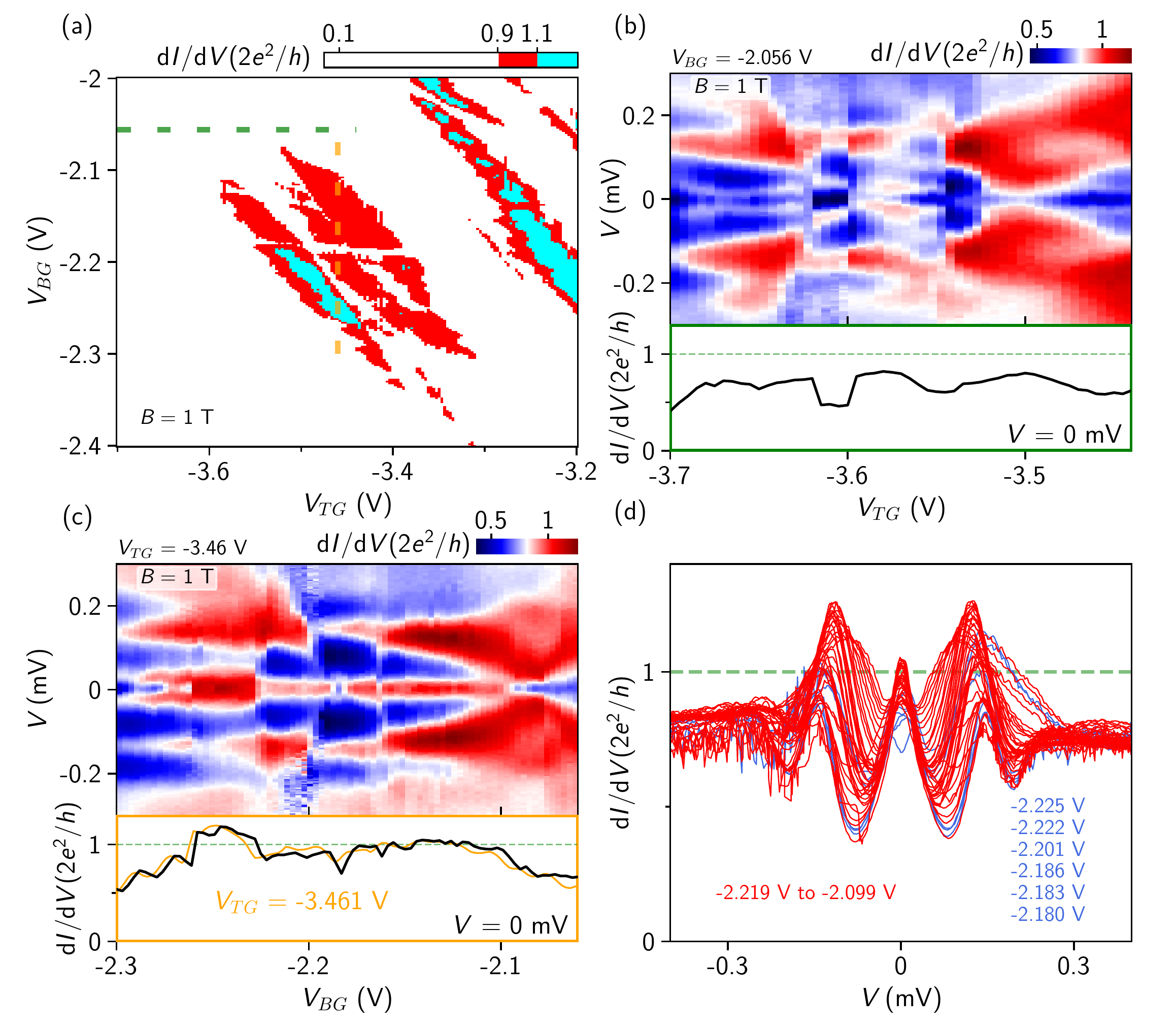}
\caption{(a) Three-color re-plot of Fig. 6b, highlighting the conductance region near $2e^2/h$ (red) with $\pm 10\%$ variation. (b) $V_{TG}$-scan at $B$ = 1 T, corresponding to the green dashed horizontal line-cut in (a), outside the `red islands'. Lower panel, zero-bias line-cut. (c) $V_{BG}$-scan at $B$ = 1 T, corresponding to the orange dashed vertical line-cut in (a). Lower panel, zero-bias line-cut (black) from the upper panel together with the vertical dashed line-cut from (a) (orange), matching qualitatively. (d) $dI/dV$ line-cuts from (c) within the $V_{BG}$-range from -2.225 V to -2.099 V. Split peaks ($V_{BG}$ labeled) and neighboring line-cuts are shown in blue for clarity.  }
\label{fig7}
\end{figure}

\begin{figure*}[htb]
\centering
\includegraphics[width=0.9\textwidth]{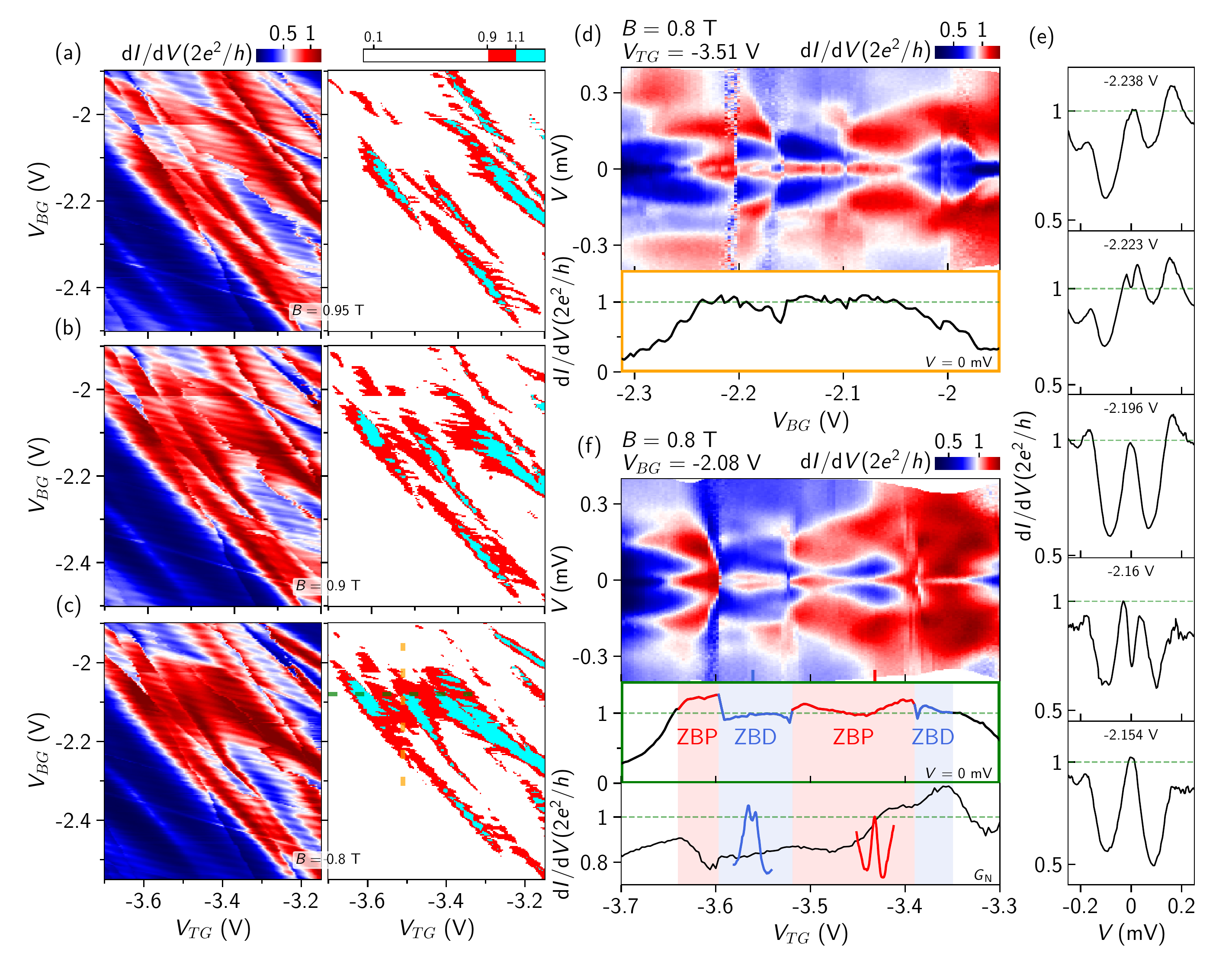}
\caption{(a,b,c) $dI/dV$ at zero bias as a function of $V_{BG}$ and $V_{TG}$ at $B$ = 0.95 T, 0.9 T and 0.8T, respectively. Right panels, three-color re-plots. (d) $V_{BG}$-scan at 0.8 T with $V_{TG}$ = -3.51 V, corresponding to the orange dashed line-cut in (c). Lower panel, zero-bias line-cut. (e) $dI/dV$ line-cuts from (d), showing the back-and-forth oscillations between ZBP and split peaks ($V_{BG}$ labeled). (f) $V_{TG}$-scan at 0.8 T with $V_{BG}$ = -2.08 V, corresponding to the green dashed line-cut in (c). Middle panel, zero-bias line-cut (red and blue for ZBP and ZBD regions). Lower panel, $G_N$: the average of conductance at the most positive and negative bias available in the data. Red and blue curves are two vertical line-cuts from the upper panel (labeled by the corresponding color bars), resolving a ZBP and a ZBD near $2e^2/h$ (horizontal ($V$) axis not shown). }
\label{fig8}
\end{figure*}

\begin{figure}[htb]
\centering
\includegraphics[width=\columnwidth]{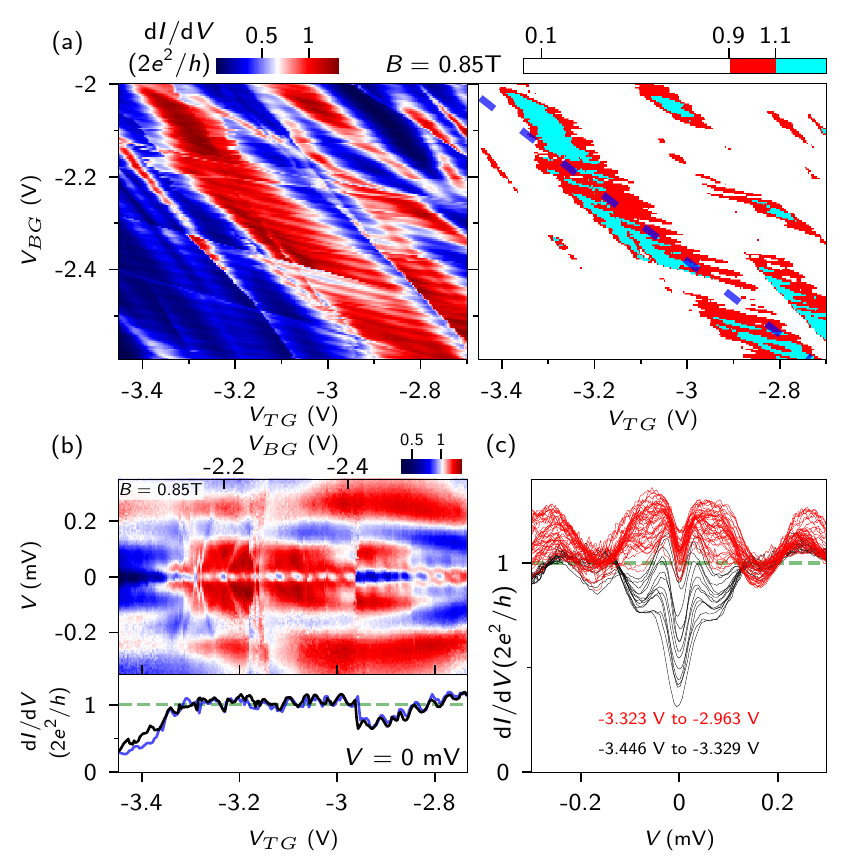}
\caption{(a) $dI/dV$ versus $V_{BG}$ and $V_{TG}$ at $B$ = 0.85 T with $V$ = 0. Right panel, three-color re-plot. (b) $dI/dV$ versus $V$ by sweeping $V_{TG}$ and $V_{BG}$ simultaneously (following the blue dashed line in (a)). Lower panel, zero-bias line-cut (black) and the blue dashed line-cut (blue) from (a). (c) Line-cuts from (b) within the $V_{TG}$-range from -3.446 V to -2.963 V (see labeling). For clarity, only one third of the line-cuts are plotted (one for every three neighboring curves). }
\label{fig9}
\end{figure}

MZM theory predicts a quantized zero-bias conductance at zero temperature. Whether this resolves a ZBP or ZBD depends on the barrier height. For a high barrier (therefore low transmission, tunneling regime), the sub-gap conductance is suppressed and a quantized ZBP can be resolved. For a low barrier with high total transmission (e.g. more than one sub-band occupation in the barrier), the sub-gap conductance can be enhanced due to Andreev reflection and exceeds $2e^2/h$. Meanwhile, the zero-bias conductance, due to spin-filtering of MZMs, still stays at $2e^2/h$ and a quantized ZBD can be resolved \cite{Wimmer2011QPC, WimmerQuasi, NextSteps}. Theory has proposed using the quantized ZBP-ZBD transition, enabled by lowering the barrier height, as an experimental tool to exclude a case of quasi-MZMs from topological MZMs \cite{WimmerQuasi, NextSteps}. In realistic situations, exact MZM quantization is not expected due to finite temperature and wire length. For a quantized ZBP with FWHM of 0.1 mV, thermal broadening of 50 mK can already decrease the peak height by $\sim$ 2$\%$ of $2e^2/h$. This height will decrease more for narrower ZBPs. To minimize temperature effect, searching for quantized MZM peak requires a large peak width, thus large barrier transmission (high $G_N$). In fact, $G_N$ of the ZBPs in Fig. 2c is close to $2e^2/h$, suggesting that the barrier is in open regime (instead of tunneling) with more than one spin-resolved sub-band occupied. In this open regime, the sub-gap conductance is finite due to Andreev reflection \cite{BTK}, resulting in a sizable background conductance for the ZBPs. This finite (Andreev) sub-gap conductance, superficially similar to a soft gap, does not ‘hurt’ the quantized height of MZM peak \cite{DasSarma2017QZBP}, different from the true soft gap case which destroys the quantization due to dissipation broadening \cite{LiuDissipation}. To confirm its 'hard gap' nature, we tune $V_{TG}$ into tunneling regime where $dI/dV$ indeed resolves a hard gap at zero $B$ (Fig. 1c) and finite $B$ (SFig. 2) for $V_{BG}$ = -2.251 V, same with the $V_{BG}$-value in Fig. 2. Overall, our ZBP is large and different from the ZBPs with small net height \cite{Yu} because 1) the zero-bias conductance is close to $2e^2/h$, and 2) the ZBP’s net height (above background) is also large and can exceed $e^2/h$. Another non-negligible effect in realistic devices is the finite wire length. In fact, our device has a relatively short length $\sim$ 658 nm for the superconducting part (Fig. 2b). As a result, MZMs could easily overlap which can further degrade the quantization quality \cite{DasSarma2012Splitting}.  

The discussion above serves as a background introduction and by no means suggests our observation as topological MZMs. In Fig. 3 we study an alternative explanation involving quasi-MZMs, solely for a qualitative illustration purpose rather than quantitative comparisons since many experimental parameters are unknown. Considering the nanowire tapering, we assume a tilted potential landscape shown in Fig. 3a (black curve). We note this is not the only possible landscape, e.g. SFig. 6 shows another case (triangle) which gives similar results. Fig. 3b shows the energy spectrum where the zero-energy state at finite Zeeman energy $V_z$ is a pair of quasi-MZMs. Fig. 3a plots the wavefunctions of these two quasi-MZMs (red $\gamma_1$ and blue $\gamma_2$), located where the chemical potential crosses the Zeeman split potential landscapes (red and blue lines). Though $\gamma_1$ and $\gamma_2$ are spatially separated, they are not at the wire ends, thus topologically trivial. Due to the separation, $\gamma_1$ has a much stronger coupling ($\Gamma_1$) to the probe than $\gamma_2$ ($\Gamma_2$, almost negligible), leading to quantized conductance as shown in Fig. 3cd. For $dI/dV$ calculation, we assume a very narrow and high barrier (vertical black line in Fig. 3a). We further assume that the barrier height is $V_z$-dependent (Fig. 3c upper): slightly decreases first, then increases, trying to capture the $B$-dependent FWHM of ZBPs and the gap shape at 0 T in Fig. 2. The physics mechanism of this assumption is not fully clear and might possibly be related to $B$-induced suppression of Andreev reflection or shifting of dot levels, both affecting the barrier transmission. With this barrier assumption, we find the numerical simulation (Fig. 3cd) qualitatively consistent with our observation of dip-to-peak transition near $2e^2/h$.

We note the interpretation above is not exclusive, e.g. landscapes with various disorder which have been extensively studied before \cite{GoodBadUgly} can not be ruled out at this stage. 

Fig. 4a-f show similar $B$-scans at three different gate voltage settings where the zero-bias conductance stays close to $2e^2/h$, persisting over sizable $B$-ranges: 0.54 T, 0.55 T and 0.56 T, respectively. The mean and standard deviation of zero-bias conductance within these $B$-ranges are, 0.95 $\pm$ 0.02, 1.01 $\pm$ 0.04 and 1.02 $\pm$ 0.04, respectively (all in unit of $2e^2/h$). The blue and red line-cuts indicate the large ZBDs and ZBPs close to $2e^2/h$. The relative variation of ZBP-width for the red line-cuts in Fig. 4df is significantly larger than the relative variation of ZBP-height near $2e^2/h$, similar to Fig. 2 (Fig. 4b has too few data points to conclude). For higher $B$, the ZBP height decreases away from $2e^2/h$, accompanied by an increase of FWHM in roughly similar $B$-ranges (see Fig. 4ce), possibly due to peak splitting. See SFig. 5 line-cuts.

In Fig. 2 and Fig. 4a-f, we have shown smooth ZBD-ZBP transitions near $2e^2/h$ with small fluctuations, forming `plateau-like' features. Based on measurements performed so far, we have not observed similar behavior at other values significantly different from $2e^2/h$. This possibility, however, could not be completely excluded since we did not (also can not) exhaust the entire ($V$, $B$, $V_{TG}$, $V_{BG}$) multi-dimensional parameter space.

In Fig. 4g, h and i, we show three more $B$-scans where the maximum ZBP-heights are (g) close to, (h) above and (i) below $2e^2/h$. We note that there is no clear boundary between Fig. 4g-i and the `plateau-like' features. Instead, we expect a smooth crossover between these behaviors tuned by gate voltages. For example, if the zero-bias conductance in Fig. 4i (Fig. 4c) was higher and more (less) flat by tuning gate, it may evolve to Fig. 2 (Fig. 4h). If the $B$-range in Fig. 4e was narrower, it may evolve to Fig. 4g. Though this smooth transition may be expected in quasi-MZMs: tuning gate voltages or $B$ may affect $\Gamma_2$ and $E_M$ (coupling between $\gamma_1$ and $\gamma_2$) which cause deviations from $2e^2/h$, below or above, both possible \cite{WimmerQuasi, JouriThesis}. The ZBPs exceeding $2e^2/h$ at many different gate voltages also strongly suggest the presence of disorder \cite{DasSarma2021above} whose detailed simulation is beyond the scope of this paper. For a full overview, we show ten more $B$-scans in SFig. 7.

Next we study $B$-scans by fixing bias to zero, see Fig. 5ad the zero-bias conductance map as a function of $B$ and $V_{BG}$ (Fig. 5a) or $V_{TG}$ (Fig. 5d). Right panels are re-plots using only three colors to highlight the conductance regions (red) close to $2e^2/h$ within $\pm10\%$ variation. We note that this range of 10$\%$ is subjective: smaller variations surely lead to smaller areas of `red islands' in three-color re-plots. The red curves in Fig. 5be are horizontal and vertical line-cuts across the ‘red islands’, resolving ‘plateau-like’ features. The $B$-scan `plateaus' have gate voltage settings close to Fig. 4ae whose zero-bias line-cuts show similar match. The $V_{BG}$-scan `plateau' at 0.84 T (lower panel b) is resolved as large ZBPs in further bias scan at a lower $B$ (0.8 T) and slightly different $V_{TG}$ (Fig. 8d). The $V_{TG}$-scan at 0.98 T (lower panel e) shows sizable fluctuations near $2e^2/h$. Further bias dependence of this curve at similar $B$ (1 T) also resolves large ZBPs (Fig. 6d). For comparisons, the black curves in Fig. 5be are line-cuts not passing through the ‘red islands’. See SFig. 8 more line-cuts of Fig. 5ad. 

The zero-bias maps only serve as a guidance in ZBP searching but does not guarantee `it is a ZBP when sweeping bias'. For example, the two black curves in Fig. 5b show a ‘peak’ above $2e^2/h$ in $B$- and $V_{BG}$-scans, corresponding to the cyan region in the lower left part of Fig. 5a (right panel). This ‘peak’ turns out not being a ZBP in further bias scan as shown in Fig. 5c. Another example is shown in Fig. 5f, a horizontal line-cut from Fig. 5d, resolving a `plateau' feature in $B$-scan at a non-quantized value of $\sim$ 0.3×$2e^2/h$. Further bias scan (Fig. 5g) on this ‘plateau’ reveals split peaks instead of ZBPs (line-cuts taken from Fig. 6d). 

Now we fix $B$ and study the gate dependence of the large ZBPs. Fig. 6a shows the zero-bias conductance map as a function of $V_{BG}$ and $V_{TG}$ at $B$ = 0 T. We note the $V_{BG}$-values and $V_{TG}$-values for the $B$-scans of ZBPs, i.e. Fig. 2, Fig, 4 and SFig. 7 (except for the two lower right panels), are all within the scanned $V_{BG}$-range and $V_{TG}$-range as shown in Fig. 6ab. The several `red-line' features are likely due to states of unintentional quantum dots formed near the barrier. A horizontal line-cut (lower panel) resolves these dot states or levels as peaks in $V_{TG}$-scan. Fig. 6c further shows the bias dependence of this line-cut where these dot states do not reveal clear and robust ZBPs (see line-cuts in Fig. 6e). At $B$ = 1 T, in addition to the dot states as a background, continuous ‘red islands’ of conductance near $2e^2/h$ are observed, see Fig. 6b whose $V_{BG}$-range and $V_{TG}$-range are almost the same as Fig. 6a. Bias scan across this ‘island’ resolves large ZBPs whose height oscillates around $2e^2/h$ (Fig. 6d), accompanied by peak-splittings. Line-cuts of the ZBPs and split peaks are shown in Fig. 6f with corresponding $V_{TG}$-ranges labeled. We note that Fig. 6bd were measured under nominally the same gate voltage settings ($V_{TG}$-range slightly different) with Fig. 6ac but at 1 T. In addition, the lower panels of Fig. 6abcd (zero-bias line-cuts) were measured at the same $V_{BG}$-value. Therefore, comparing these four line-cuts allows to identify the sharp `jump-like' features and to what extent they can be reproduced upon re-measuring. For example, most of the oscillating features near $2e^2/h$ in the lower panel of Fig. 6d are reproducible based on the comparison with the lower panel of Fig. 6b where matches can be found. These features, despite being reminiscent of charge jumps, are reproducible sharp resonances (possibly dot levels) tuned by gate voltages. The left most ‘jump’ in Fig. 6d ($V_{TG}$ $\sim$ -3.57 V) is a non-reproducible charge jump which is absent in Fig. 6b. 

Fig. 7a shows the three-color re-plot of Fig. 6b, highlighting the ‘red islands’ as regions close to $2e^2/h$. In Fig. 6d we have studied a horizontal line-cut across the ‘red islands’, showing large ZBPs. For comparison, Fig. 7b shows bias dependence of a  horizontal line-cut outside the ‘red islands’ (green dashed line in Fig. 7a). No clear and robust ZBPs are observed (occasionally there are non-robust ZBPs due to sharp level crossing like SFig. 4bc). Next, in Fig. 7c, we study a vertical line-cut across the ‘red islands’ (orange dashed line in Fig. 7a) where the bias dependence also resolves large ZBPs. These large ZBPs show a ‘plateau-like’ feature near $2e^2/h$ in zero-bias conductance (lower panel) for $V_{BG}$ from -2.225 V to -2.099 V. All the $dI/dV$ line-cuts within this quoted $V_{BG}$-range are shown in Fig. 7d (split peaks and their neighboring line-cuts in blue for clarity). The mean and standard deviation of the zero-bias conductance for these line-cuts (including the split peaks) in Fig. 7d is 0.95 $\pm$ 0.07 in unit of $2e^2/h$. The split peaks may be due to non-negligible coupling ($E_M$) between quasi-MZMs at that particular gate voltage setting. For lower $V_{BG}$-values ($\sim$ -2.25 V), the ZBP-height significantly exceeds $2e^2/h$ (maximally $\sim$ 1.2$\times 2e^2/h$), possibly due to non-negligible coupling of the second quasi-MZM to the barrier ($\Gamma_2$) enabled by smooth potential variation or purely a disorder effect \cite{DasSarma2021above}. See SFig. 9 for more $V_{BG}$- and $V_{TG}$- scans of the large ZBPs at this field of 1 T and SFig. 10 more line-cuts from Fig. 6b. 

After the extensive gate scans at $B$ = 1 T, we now tune $B$ to other values of 0.95 T, 0.9 T and 0.8 T, respectively, as shown in Fig. 8a-c. The right panels are the three-color re-plots with ‘red islands’ highlighting regions close to $2e^2/h$. Comparing Fig. 8a-c with Fig. 6b, we can find matches for the main features (dot states or resonances) with minor overall gate voltage drifts, possibly due to small charge jumps happened between those measurements which reset/shift the overall gate voltage. Fig. 8d shows bias dependence for a vertical line-cut across the `red islands' in Fig. 8c (orange dashed line), resolving large ZBPs and split peaks near $2e^2/h$. The zero-bias conductance of these ZBPs and split peaks form a `plateau-like' feature with conductance fluctuating around $2e^2/h$ for $V_{BG}$ between -2.238 V and -2.046 V. The mean and standard deviation of the zero-bias conductance within this $V_{BG}$-range, including both ZBPs and split peaks, is 1.00 $\pm$ 0.08 in unit of $2e^2/h$. Fig. 8e shows several $dI/dV$ line-traces (with $V_{BG}$-values labeled) from Fig. 8d, illustrating the back-and-forth oscillating behavior for ZBPs vs splitting peaks. 

Fig. 8f shows a $V_{TG}$-scan of the ZBPs, corresponding to a horizontal line-cut from Fig. 8c (green dashed line). The red and blue colors indicate the $B$-ranges for ZBPs and ZBDs, which alternate for a total $V_{TG}$-range from -3.642 V to -3.342 V. Within this range, the mean and standard deviation of zero-bias conductance is 1.06 $\pm$ 0.09 in unit of $2e^2/h$. The transitions between ZBP and ZBD regions are accompanied by sharp and reproducible resonances, possibly due to dot levels crossing zero energy. These dot levels, tuned by gate voltages, can interact with quasi-MZMs when approaching zero energy. Theory has suggested that the energy hybridization between these dot levels and the zero-energy states can reveal asymmetric couplings of the two quasi-MZM components to the normal probe \cite{Prada2017, Clarke2017}. Fig. 8f further shows that this interaction could also interfere with the zero-bias conductance, causing sizable deviations/fluctuations from/around $2e^2/h$. All the $dI/dV$ line-cuts within the $V_{BG}$ and $V_{TG}$ ranges mentioned above can be found in SFig. 11. The fluctuations of zero-bias conductance near $2e^2/h$ and the alternating peak-split peak behavior shown in Fig. 8df could possibly be related to an oscillatory couplings ($E_M$) between the two quasi-MZMs and the coupling of the second quasi-MZM to the barrier ($\Gamma_2$), both tuned by gate voltages in a smooth or disordered potential landscape. For more vertical and horizontal line-cuts at 0.8 T (Fig. 8c) as well as 1.0 T (Fig. 6b), see SFig. 10.

So far, we have presented several quantized `plateau-like' features for the zero-bias conductance in $B$-scan (Fig. 2, Fig. 4a-f) and gate voltage scans (Fig. 7c, Fig. 8d), as well as several other non-plateau scans at different $B$ and gate voltage settings. These `plateau-like' features have noticeable fluctuations: some smaller and some larger, but generally within $\pm$10$\%$ variation of the quantized value. Though perfect quantization of MZM is also not expected in realistic devices with finite temperature and short wire length as discussed before, we believe there is still much room for improvement regarding the flatness and accuracy of plateaus based on our current data quality. We further note that plateau-like features and non-plateau features are not `black vs white' with clear and sharp boundaries, as was partially discussed before in Fig. 4. For example, our $V_{TG}$-scans (Fig. 6d, Fig. 8f) in general show larger fluctuation amplitudes which probably can not be identified as `plateau-like'. But they do show large ZBPs oscillating around $2e^2/h$, different from those non-plateau features: Fig. 7b with no robust ZBPs and SFig. 4bc with small ZBPs due to sharp level crossings (thus non-robust). Therefore, these $V_{TG}$-scans (Fig. 6d, Fig. 8f) can be treated as intermediate cases or transitions between plateau-like and non-plateau features. With future device optimization, e.g. reducing disorder, these features may develop into plateau-like or plateau features \cite{Tudor2021Disorder}.

Finally, we show a gate scan at $B$ = 0.85 T where we find ZBDs as the dominating feature. Fig. 9a shows the zero-bias conductance map as a function of $V_{BG}$ and $V_{TG}$. We notice a significant charge jump between the measurement of Fig. 9 and the rest majority of the data (Fig. 2 to Fig. 8). As a result, the main features (dot states) of the zero-bias map at different fields (Fig. 6b and Fig. 8a-c) do not show clear match with Fig. 9a. Therefore, the data set of Fig. 9 is isolated and can not be compared directly with the rest (Fig. 2 to Fig.8). Nevertheless, we can still resolve ‘red islands’ as shown in the right panel. The blue dashed line (tuning $V_{BG}$ and $V_{TG}$ simultaneously) marks a fine-tuned cut passing through the ‘red-islands’ which resolves a `plateau-like' feature around $2e^2/h$ for $V_{TG}$ from -3.323 V to -2.963 V, as shown in Fig. 9b. The mean and standard deviation of the zero-bias conductance within this gate range is 1.05 $\pm$ 0.06 in unit of $2e^2/h$. $dI/dV$ line-cuts (Fig. 9c) within the `plateau-like' region resolve zero-bias dips (ZBD) with no clear and robust ZBPs. For quantized ZBDs due to quasi-MZMs or MZMs, thermal averaging effect at finite temperature tends to increase the zero-bias conductance above $2e^2/h$, contrary to the case of quantized ZBPs. We further note the small quasi-periodic oscillations for gate sweeps in Fig. 9b and Fig. 9a. Occasionally, they resolve diamond-like shapes in bias vs gate scan as shown in Fig. 9b with a diamond size of $\sim$ 0.2 mV or smaller, similar to Coulomb blockades. These small quasi-periodic oscillations can also be found in Fig. 6ab and Fig. 8a-d. We do not know the origin of these oscillations but speculate that it may be related to the short length of the nanowire (the superconducting part $\sim$ 658 nm) which might be treated as an open and large quantum dot with a small charging energy.  \newline

\section{Summary}

To summarize, we have measured large zero-bias peaks on the order of $2e^2/h$ in a thin InAs-Al hybrid nanowire device, using a four-terminal device design. At particular gate voltage settings, we observe a smooth transition between zero-bias peaks and zero-bias dips, driven by magnetic field. The zero-bias conductance sticks close to $2e^2/h$ during this dip-to-peak transition, forming a plateau-like feature. Further gate scans of these zero-bias peaks at finite magnetic field reveal plateau-like features (with fluctuations) around $2e^2/h$. We discuss our data with a possible (not necessary exclusive) interpretation based on quasi-Majorana zero modes, smooth potential variation and disorder. All the results in this paper are from a single-device, and future devices could be aiming at longer and more uniform (non-tapered) thin InAs-Al wires with better gate and dielectric designs to minimize the level of disorder, which hopefully may lead to better quantization: more flat and accurate plateaus.

\begin{acknowledgments}
We thank Leo Kouwenhoven, Sankar Das Sarma, {\"O}nder G{\"u}l and Guanzhong Wang for valuable comments and a critical reading of the manuscript. Raw data and processing codes within this paper are available at http://doi.org/10.5281/zenodo.5111639. This work is supported by Tsinghua University Initiative Scientific Research Program, National Natural Science Foundation of China (Grant Nos. 92065106 $\&$ 61974138), Beijing Natural Science Foundation (Grant No. 1192017). H. S. acknowledges China Postdoctoral Science Foundation (2020M670173 $\&$ 2020T130058), D. P. also acknowledges the support from Youth Innovation Promotion Association, Chinese Academy of Sciences (No. 2017156).
\end{acknowledgments}

\bibliography{mybibfile}% Produces the bibliography via BibTeX.

\newpage
\onecolumngrid


\section*{Supplementary material}

# Supplement Information: Large zero bias peaks and dips in a four-terminal thin InAs-Al nanowire device

Huading Song,[1,2,*] Zitong Zhang,[1,*] Dong Pan,[2,3,*] Donghao Liu,[1] Zhaoyu Wang,[1] Zhan Cao,[2] Lei Liu,[3] Lianjun Wen,[3] Dunyuan Liao,[3] Ran Zhuo,[3] Dong E Liu,[1,2,4] Runan Shang,[2,†] Jianhua Zhao,[2,3,‡] and Hao Zhang[1,2,4,§]

[1]*State Key Laboratory of Low Dimensional Quantum Physics,
Department of Physics, Tsinghua University, Beijing 100084, China*
[2]*Beijing Academy of Quantum Information Sciences, 100193 Beijing, China*
[3]*State Key Laboratory of Superlattices and Microstructures, Institute of Semiconductors,
Chinese Academy of Sciences, P. O. Box 912, Beijing 100083, China*
[4]*Frontier Science Center for Quantum Information, 100084 Beijing, China*

**Device Fabrication:** The InAs nanowires were grown by molecular-beam epitaxy followed by an in-situ growth of Al film. The nanowires were then transferred from the growth chip to a p-doped Si/SiO$_2$ substrate by wipes of clean room tissues. Part of the Al film was selectively etched using Transene Aluminum Etchant Type D at 50 °C for 10 seconds with etch windows patterned by electron-beam lithography (EBL). Electric contacts and side gates were fabricated in another round of EBL. Before metal deposition, a 40s long Argon plasma etching at a power of 50 W and pressure of 0.05 Torr was performed in the load-lock to ensure Ohmic contacts.

**Data Analysis:** For all the 2D color maps of $dI/dV$ vs $V$ and $B$ (or gate voltages), a bias offset $V_{offset}$ ($\sim 50$ uV) is estimated and subtracted from $V$, based on the symmetry of each data set which relies on the particle-hole symmetry of the superconducting gap (see data repository for $V_{offset}$ details). For the color-map plots, the $dI/dV$ vs $V$ curves are interpolated on to a regularly spaced $V$ grid. Zero-bias line-cuts in bias scans are extracted based on the conductance value whose corresponding bias $V$ is the closest to zero.

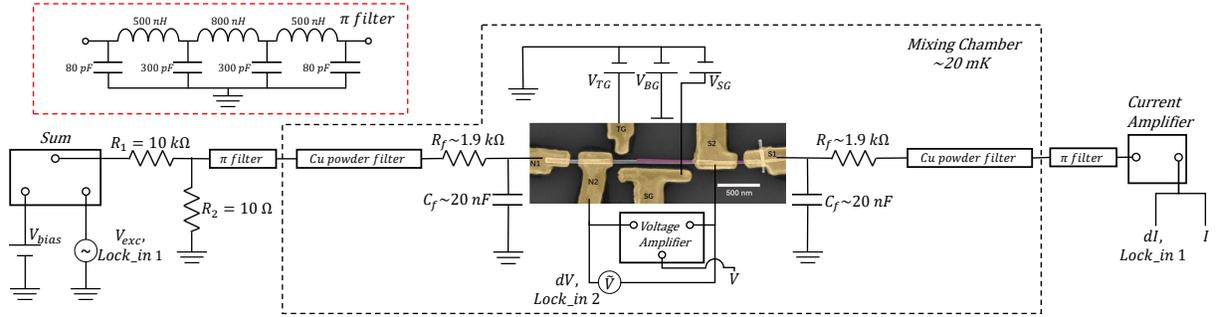

Supplement Fig. 1: Schematic of the measurement circuit. A total bias voltage $V_{bias}$ from a DC voltage source is first mixed with a lock-in excitation voltage into a summing module. This DC + AC voltage signal, after passing through a voltage divider (with $R_1 : R_2 = 1000:1$), is applied to the N1 contact with several filters in between ($\pi$-filter at room temperature, copper powder filter and RC filter in the mixing chamber). The current ($I$ and $dI$) is drained from the S1 contact, also passing through these three-stage filters, to a pre-amplifier (a current-voltage converter with an amplification gain of 1 uA/V) and measured using lock-in 1 and a DC voltage meter. In addition, the voltage drop ($V$ and $dV$) between the contacts N2 and S2 is measured using another DC voltage meter (after 100× amplification) and a second lock-in which is synchronized with the first one. Therefore, the differential conductance $dI/dV$ is directly calculated by taking $dV$ from lock-in 2, and $dI$ from lock-in 1, where $V$ can be read directly from the voltage meter between N2 and S2. We use the lock-in X-components in calculating $dI/dV$ for all data, and if using the R-components, we find $dI/dV$ to be $\sim 1\%$ higher for conductance near $2e^2/h$. Therefore, this 1% sets a lower bar of the measurement uncertainty for conductance near $2e^2/h$. Gate voltages provided by three voltage sources are applied to corresponding gates, after passing through the three-stage filters (not shown in the figure). The mixing chamber of the dilution fridge has a base temperature $\sim 20$ mK. Between the room temperature $\pi$-filters and mixing chamber filters, the fridge lines also have proper thermal anchoring at every stage.

---

[*] equal contribution
[†] shangrn@baqis.ac.cn
[‡] jhzhao@semi.ac.cn
[§] hzquantum@mail.tsinghua.edu.cn

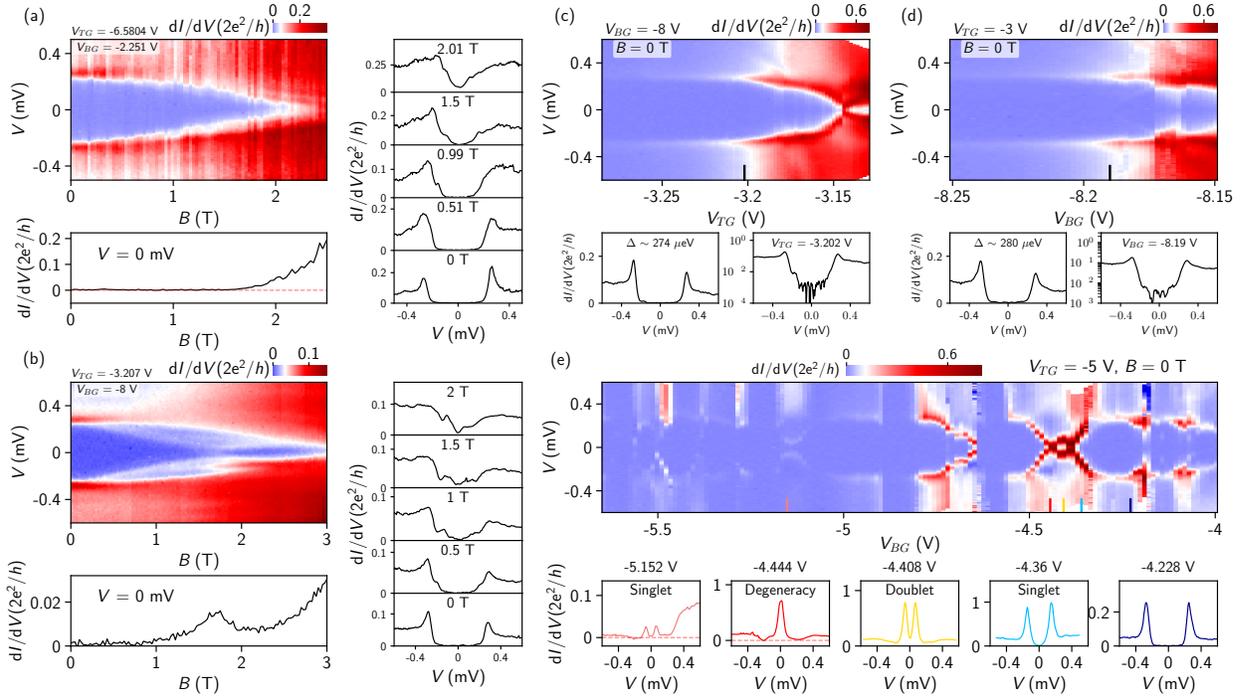

Supplement Fig. 2: Device basic characterization. (a) Magnetic field dependence of the hard superconducting gap at $V_{BG} = -2.251$ V (same with Fig. 1c and Fig. 2). Lower panel, zero-bias line-cut. Right panels, $dI/dV$ line-cuts at different magnetic fields. The gap remains hard (zero-bias conductance $\sim 0$) at finite $B$ (for $B < 1.5$ T). Note that this $B$-field range is also where we observe large zero-bias peaks and dips. (b) Same with (a) but at $V_{TG} = -3.207$ V which is close to the $V_{TG}$-value of Fig. 2. Two sub-gap states detach from the gap edges and anti-cross at $\sim 1.7$ T, resulting in an increase of the zero-bias conductance for $B > 1$ T. Our large ZBPs were measured in open regime (finite sub-gap conductance). To reach tunneling regime, one needs to decrease either $V_{TG}$ or $V_{BG}$, as was done in (a) or (b), both showing hard gaps. (c) $V_{TG}$-scan of the superconducting gap for $V_{BG} = -8$ V, resolving a hard gap (see lower panel the line-cut). This $V_{TG}$-range is close to $V_{TG}$-values of the large ZBPs. (d) $V_{BG}$-scan of the superconducting gap for $V_{TG} = -3$ V. (e) $V_{BG}$-scan over a large voltage range resolves Andreev bound states (ABSs) due to unintentional quantum dots. Lower panels show line-cuts of different ABS cases (singlet, doublet and degeneracy) with $V_{BG}$-values labeled above (also indicated by the corresponding color bars in the upper panel).

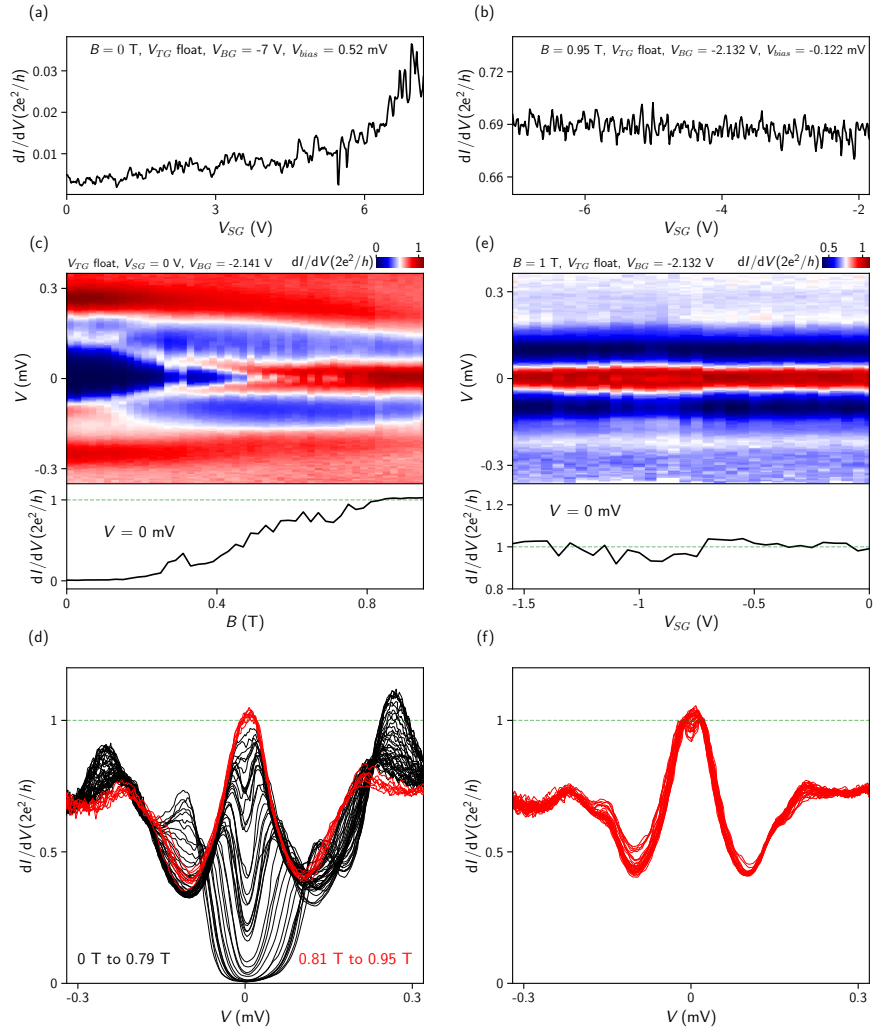

Supplement Fig. 3: Effect of super gate. (a, b) Super gate voltage ($V_{SG}$) scan near pinch-off (a) and in open regime (b) with parameters of $B$, bias and gate voltages labeled. $V_{bias}$ is the total bias applied to N1 (see SFig. 1) where, due to offsets from the circuit, $dI/dV$ at $V_{bias}$ of 0.52 mV and -0.122 mV roughly correspond to $G_N$ and zero-bias conductance, respectively. Tunnel gate was unintentionally floated due to a bad connection, realized after the measurement of this figure (and fixed). The conductance shows little response for $V_{SG}$-scan over large gate voltage ranges. (c) $dI/dV$ versus $V$ and $B$, resolving large ZBPs near $2e^2/h$. Lower panel shows the zero-bias line-cut. (d) Waterfall plot of all the $dI/dV$ line-traces from (c), red for large ZBPs near $2e^2/h$. (e) $V_{SG}$-scan of the large ZBP at $B = 1$ T, over a large gate voltage range with the zero-bias line-cut shown in the lower panel. (f) All $dI/dV$ line-traces from (e). Based on the unusual $V_{SG}$ dependence of the device conductance and ZBPs over large super gate voltage ranges (several times larger than the $V_{TG}$- and $V_{BG}$-ranges shown in the paper), we concluded that the super gate was not well functional and thus set to 0 V for the rest of the measurements (other figures within this paper). For this figure, due to noise fluctuations (likely because of the floating of tunnel gate), the $dI/dV$ line-traces show unstable wiggles if plotted using our standard four-terminal method. To overcome this situation, the bias $V$ in (c)-(f) was calibrated using the traditional two-terminal formula of $V = V_{bias} - I \times R_{Series}$, different from all other figures where $V$ was directly taken from the voltage meter between N2 and S2. $R_{Series}$ is estimated to be 19 $k\Omega$, which shows a good match for the ZBP-width and shape by plotting together with the wiggling line-trace using four-terminal method. A minimal smoothing is performed for $I$ over three neighboring points. We note the conductance value $dI/dV$ in this figure is still using the standard four-terminal method without subtracting any series resistance like all other figures.

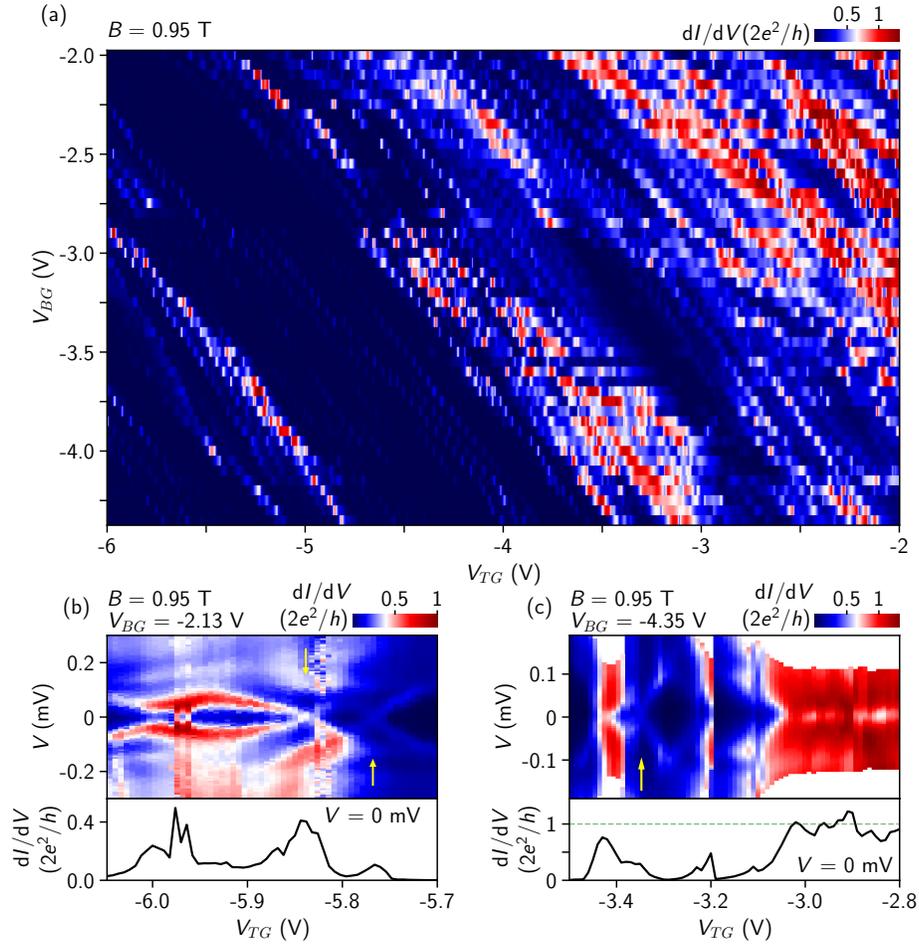

Supplement Fig. 4: Coarse search of ZBPs. (a) Zero-bias conductance map at $B = 0.95$ T. This map provides a coarse overview of the gate voltage parameter space which serves as a guidance on where to 'zoom-in' and search for large ZBPs. As shown in Fig. 5-7, large zero-bias conductance does not necessary resolve a zero-bias peak in bias sweep. (b) An unsuccessful search in the upper left region of the 'map' in (a). The zero-bias line-cut shows three peak features when sweeping $V_{TG}$. The first peak at $V_{TG} \sim$ -5.97 V is not a zero-bias peak in bias scan but split peaks. The other two peaks at $V_{TG} \sim$ -5.84 V and -5.76 V are indeed zero-bias peaks. But these two ZBPs (see yellow arrows) are 1) small in height, and 2) not robust in $V_{TG}$-scan and quickly split. These ZBPs, formed by sharp level crossings, are typical Andreev bound states. (c) Another 'zoom-in' search in the bottom part of (a), showing no robust ZBPs either. The yellow arrow marks a non-robust ZBP as a typical Andreev bound state. The large zero-bias conductance (lower panel, between -3 V and -2.8 V) does not show ZBPs in bias sweep.

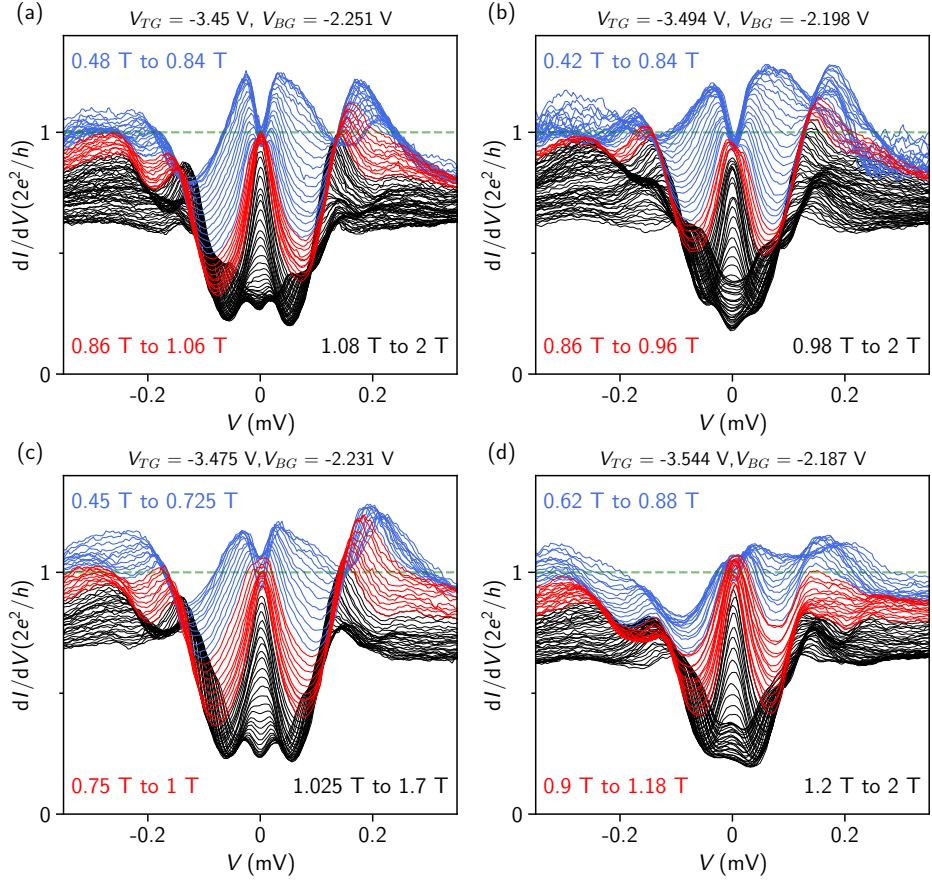

Supplement Fig. 5: Line-cuts of ZBPs and split peaks at higher $B$. (a-d) correspond to Fig. 2c, Fig. 4b, Fig. 4d and Fig. 4f, respectively. The $dI/dV$ line-traces at higher $B$-values (black curves) show a continuous decrease of zero-bias conductance which finally leads to peak splittings.

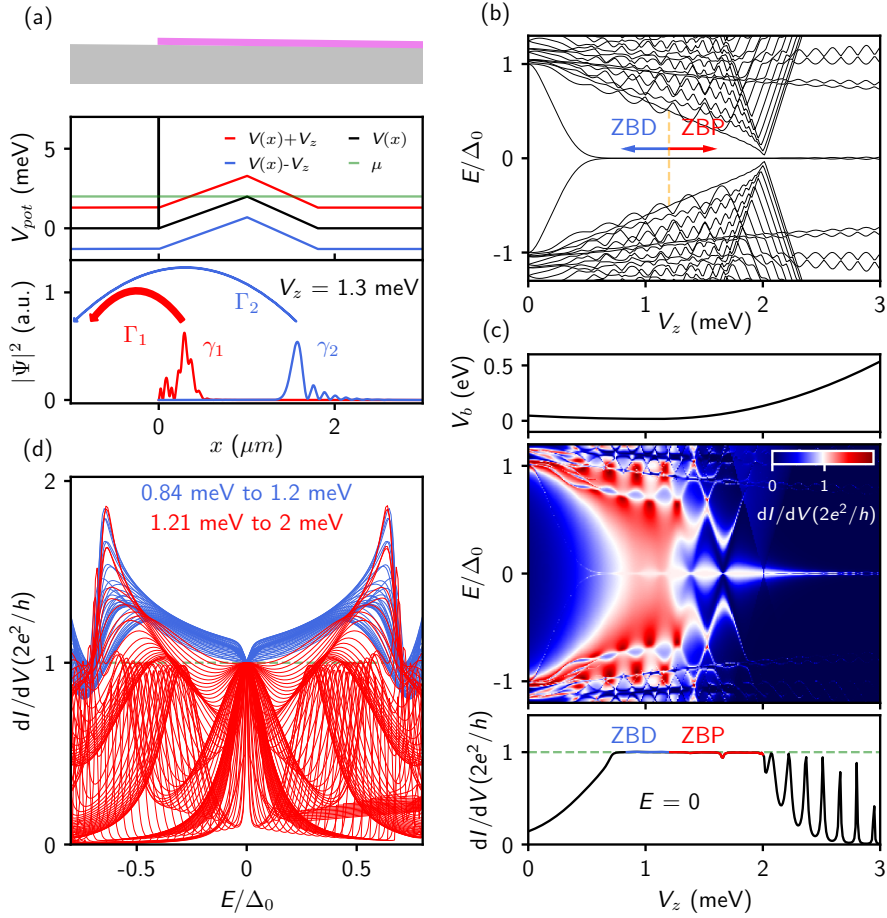

Supplement Fig. 6: More about the numerical theory model. The Bogoliubov-de Gennes (BdG) Hamiltonian of the nanowire can be written as $\mathcal{H}_{\text{NW}} = \frac{1}{2}\int dx \Psi^\dagger(x) H_{\text{NW}} \Psi(x)$ with $H_{\text{NW}}$ describing a one-dimensional semiconductor nanowire coupled to a superconductor: $H_{\text{NW}} = \left(\frac{p_x^2}{2m^*} - \mu + V(x)\right)\tau_z + \frac{\alpha}{\hbar}p_x\sigma_y\tau_z + V_z\sigma_x + \Delta_0\tau_x$. $p_x = -i\hbar\partial_x$ is the momentum, $m^*$ is the effective mass, $\mu$ is the chemical potential, $V(x)$ is the electrostatic potential along the nanowire, $\alpha$ is the spin-orbit coupling strength, $V_z$ is the Zeeman field, and $\Delta_0$ is the proximity-induced superconducting gap. $\sigma_i$ and $\tau_i$ ($i = x, y, z$) are Pauli matrices acting on spin and particle-hole space, respectively. An infinitesimal dissipation term $i\Gamma$ is also added in the Hamiltonian in order to smooth the conductance. The Nambu spinor basis is chosen as $\Psi(x) = \left(\psi_\uparrow(x), \psi_\downarrow(x), \psi_\downarrow^\dagger(x), -\psi_\uparrow^\dagger(x)\right)^T$. We chose the parameters as $m^* = 0.05$ $m_e$, $\alpha = 30$ meVnm, $\Delta_0 = 0.25$ meV, $\mu = 2$ meV. We use a triangle shape potential $V(x)$, see the black curve in (a), middle panel (in Fig. 3, the potential has a tilted line shape). The energy spectrum of this nanowire is given in (b). Below the topological phase transition point ($V_z \sim 2$ meV), a near-zero-energy state already appears as quasi-MZMs whose wavefunctions at $V_z = 1.3$ meV are shown in the lower panel of (a). Next we use a software package Kwant to calculate the conductance. We attach a lead to the left end of the nanowire and apply a bias voltage. A very narrow tunnel barrier with height $V_b$ is also added between the lead and the nanowire, as shown by the vertical line in the middle panel of (a). $V_b$ is larger than the shown range in the middle panel of (a) due to limited space (same for Fig. 3a). We further assume that $V_b$ is $V_z$-dependent which first slightly drops for a certain value and then increases with $V_z$ as shown in (c). The calculated $dI/dV$ vs bias voltage $E$ and $V_z$ is shown in (c) (middle panel). The zero-bias line-cut (lower panel (c)) shows a $2e^2/h$ quantized plateau in a $V_z$-range labeled by the blue and red lines. $dI/dV$ line-traces within this $V_z$-range are plotted in (d) showing the dip-to-peak transition. For higher $V_z$, the zero-bias conductance starts to oscillate (also shown in Fig. 3c), due to overlapping of MZMs. These ZBPs (at high $V_z$) have very narrow peak width which could easily be smoothed by thermal broadening (thus height drops below $2e^2/h$) in realistic situation with finite device temperature.

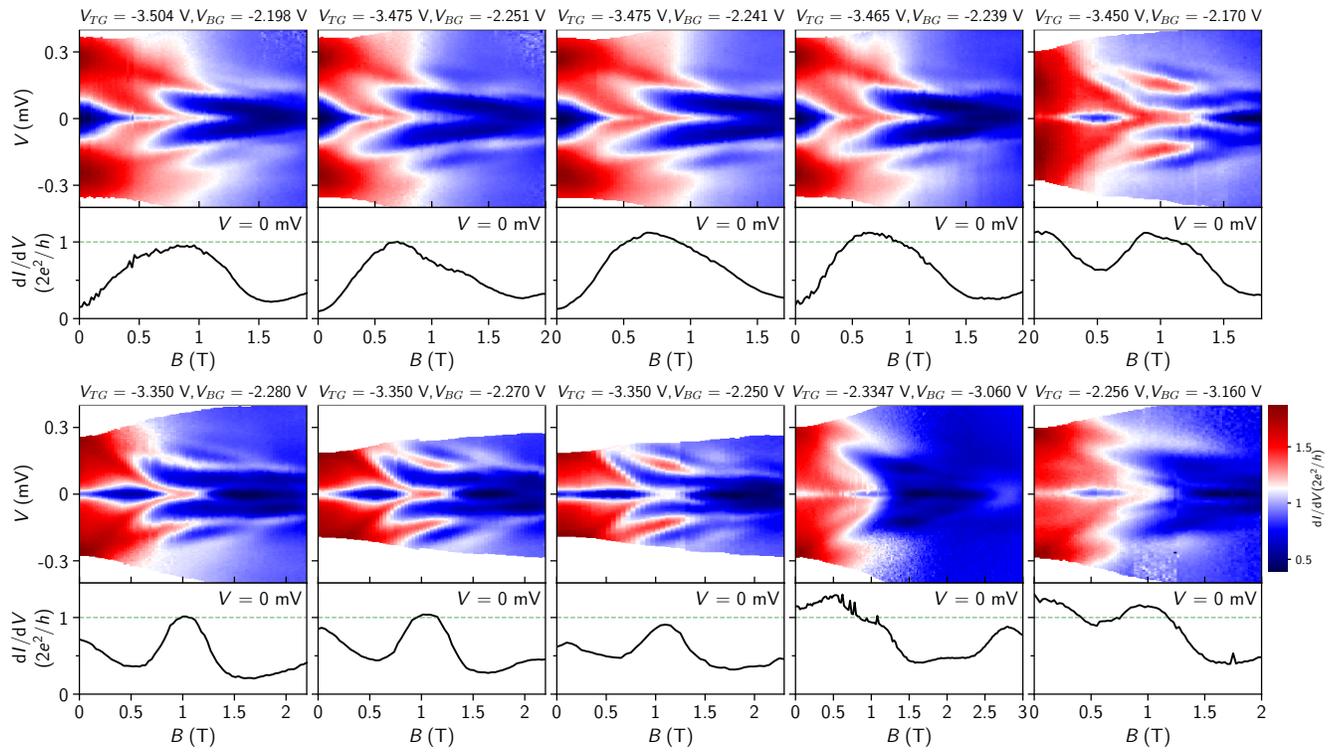

Supplement Fig. 7: More $B$-scans of ZBPs at various gate voltage settings (labeled in each panel). Lower panels show the zero-bias line-cuts.

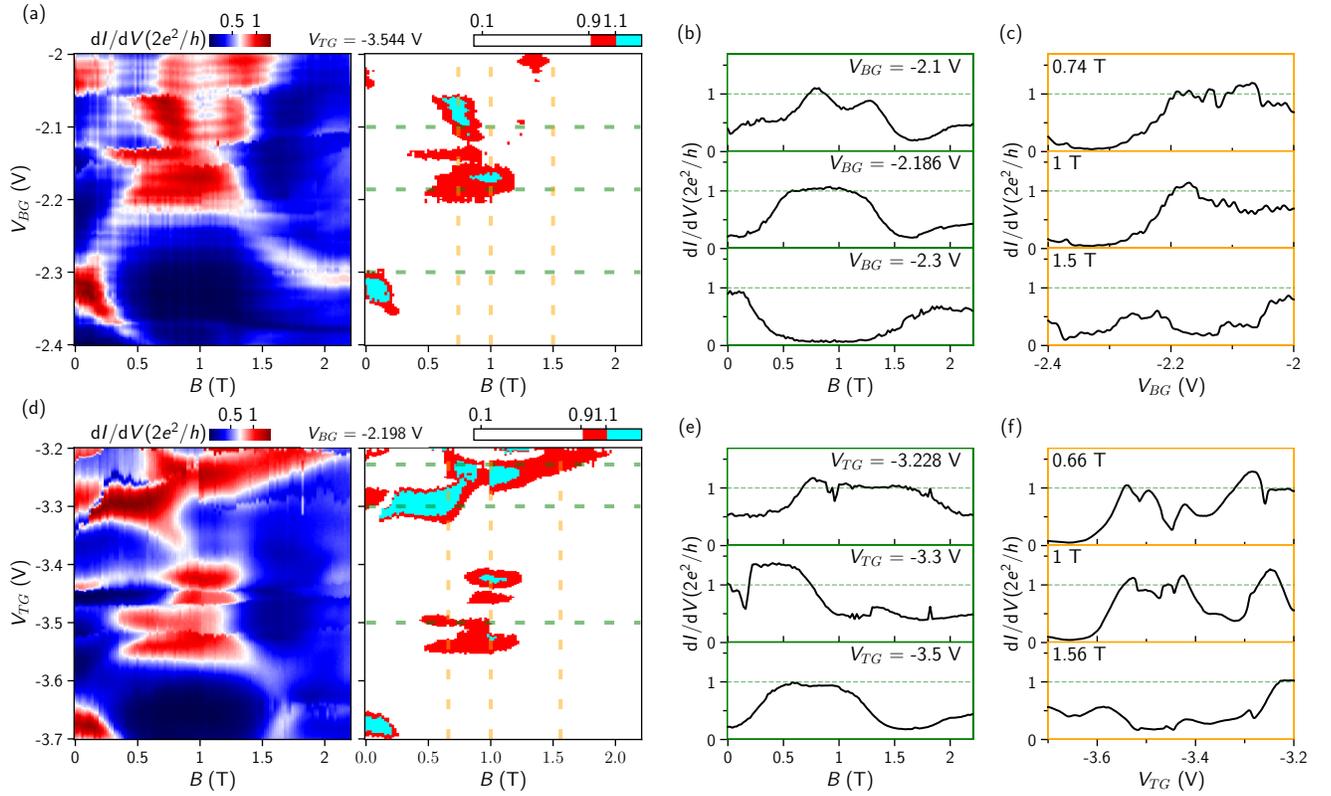

Supplement Fig. 8: More line-cuts of Fig. 5. (a) and (d), same with Fig. 5a and 5d. (b,c) and (e,f) show more horizontal/vertical line-cuts in (a) and (d), respectively. The gate voltages and magnetic fields are labeled in each panel, also indicated by the dashed lines in the right panels of (a) and (d). The $B$-scan 'plateau' in the middle panel of (b) shares almost the same gate voltage settings with Fig. 4ef whose zero-bias line-cut shows similar 'plateau' feature. The small 'plateau' at $2e^2/h$ near zero $B$ in the lower panel of (b) is likely not a ZBP based on the bias scan in Fig. 5c. Another $B$-scan 'plateau' in the lower panel of (e) have gate voltage settings very close to Fig. 4ab and SFig. 7 (the first panel). As for the 'plateau' above $2e^2/h$ in the middle panel of (e), we do not have the bias scan data near this parameter space and therefore could not identify it as ZBPs or non-ZBPs. The middle panel in (f) corresponds to Fig. 6d with the same $V_{BG}$ and $B$.

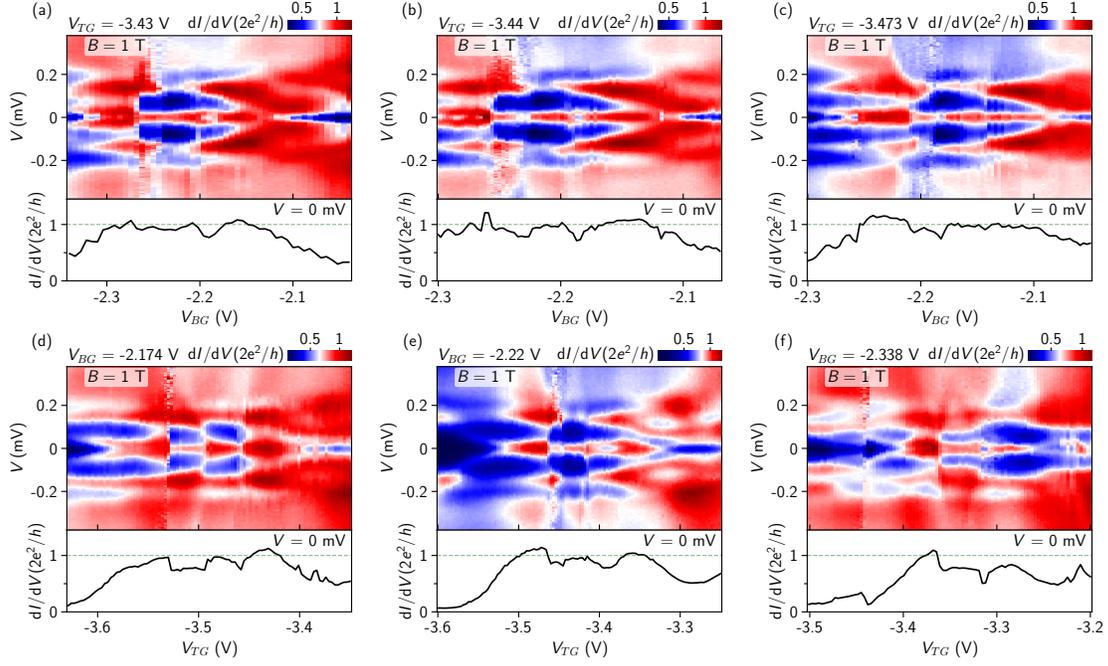

Supplement Fig. 9: More gate scans of ZBPs at $B = 1$ T. (a)-(c) $V_{BG}$-scans of the ZBP at different tunnel gate voltages labeled in each panel. Zero-bias line-cuts are shown in the lower panels. (d)-(f) $V_{TG}$-scans of ZBPs at different back gate voltages labeled in each panel. Zero-bias line-cuts are shown in the lower panels. The zero-bias line-cut in (f) may possibly be argued as a 'plateau-like' feature near $0.8 \times 2e^2/h$. Its back gate voltage corresponds to a horizontal line-cut near the edge of the 'red islands' in Fig. 7a, if assuming no gate voltage drift in between, while (d) and (e) correspond to line-cuts across the 'red islands'. Therefore, we think (f) is an intermediate case (transition or crossover) from 'plateau-like' to non-plateau situation. This also suggests that to fully establish MZM or quasi-MZM quantization with reasonably good quality, more efforts are needed, e.g. by minimizing disorder and observing more flat and accurate plateaus or providing additional experimental signatures (e.g. dip-to-peak transition).

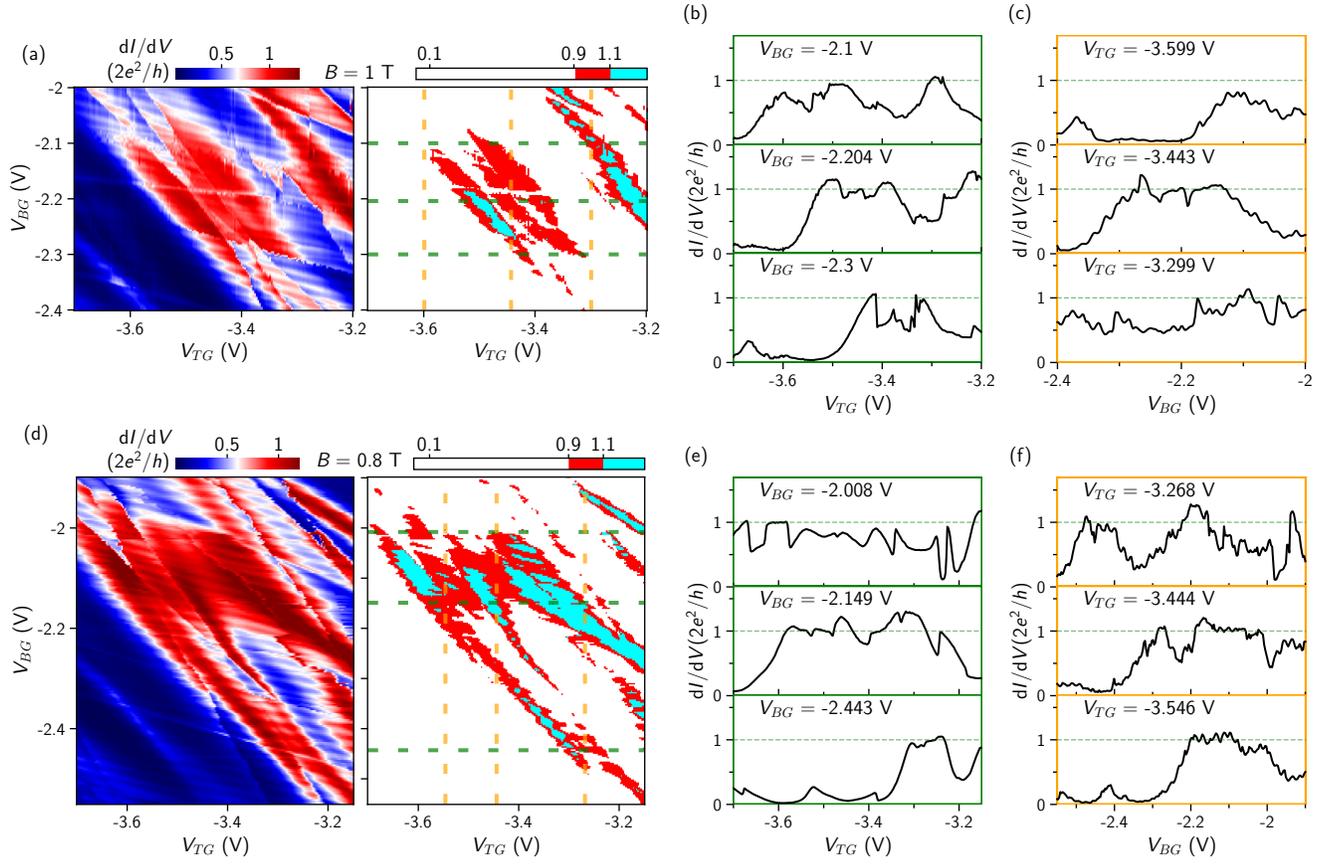

Supplement Fig. 10: More line-cuts of the zero-bias map at 1 T and 0.8 T. (a) and (d), same with Fig.6b (Fig. 7a) and Fig. 8c. (b) and (c) show the horizontal and vertical line-cuts from (a) (see corresponding dashed lines). (e) and (f) show the horizontal and vertical line-cuts from (d) (see corresponding dashed lines).

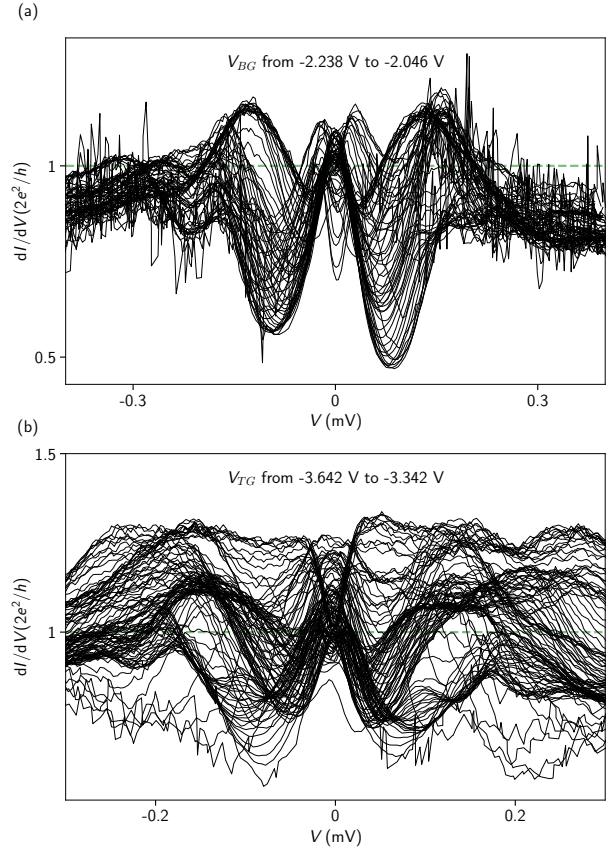

Supplement Fig. 11: (a) All the line-cuts from Fig. 8d within the $V_{BG}$ range from -2.238 V to -2.046 V. (b) All the line-cuts from Fig. 8f within the $V_{TG}$ range from -3.642 V to -3.342 V



\end{document}